\documentclass[%
reprint,
superscriptaddress,
nobibnotes,
 amsmath,amssymb,
 aps,
prb,
footinbib,
]{revtex4-2}
\usepackage{graphicx,subfigure}
\usepackage{epsfig}
\usepackage{bm}				
\usepackage{dcolumn}
\usepackage{color,xcolor}
\usepackage{physics}
\usepackage{float}
\usepackage[colorlinks,urlcolor=blue,citecolor=blue,linkcolor=magenta]{hyperref}
\makeatletter
\newcommand*{\rom}[1]{\expandafter\@slowromancap\romannumeral #1@}
\makeatother
\usepackage{lipsum}
\usepackage{subfigure}
\usepackage{dsfont}
\usepackage{enumitem}
\usepackage{tikz,verbatim}
\usetikzlibrary{quantikz}

\newcommand{\hamt}{\hat{\mathcal{H}}\left(t\right)}

\def\RBB{\textcolor{black}}

\newcommand{\tc}[1]{\textcolor[rgb]{0,0.0,0.0}{#1}}

\begin{document}

\title{A Robust Large-Period Discrete Time Crystal and its Signature in a Digital Quantum Computer}

\author{Tianqi Chen}
\email{{chen\_tianqi@a-star.edu.sg}}
\affiliation{{Bioinformatics Institute, Agency for Science, Technology and Research (A*STAR), 30 Biopolis Street, No.~07-01 Matrix, Singapore 138671}}
\affiliation{{Quantum Innovation Centre (Q.~InC), Agency for Science, Technology and Research (A*STAR), 2 Fusionopolis Way, Innovis No.~08-03, Singapore 138634}}
 \affiliation{Department of Physics, National University of Singapore, Singapore 117542}
 \affiliation{{Centre for Quantum Technologies, National University of Singapore, Singapore 117543}}
 \affiliation{School of Physical and Mathematical Sciences, Nanyang Technological University, Singapore 639798}
 \author{Ruizhe Shen}
\email{ruizhe20@u.nus.edu}
\affiliation{Department of Physics, National University of Singapore, Singapore 117542}
\author{Ching Hua Lee}
\email{phylch@nus.edu.sg}
\affiliation{Department of Physics, National University of Singapore, Singapore 117542}
\author{Bo Yang}
 \affiliation{School of Physical and Mathematical Sciences, Nanyang Technological University, Singapore 639798}
 \affiliation{Institute of High Performance Computing (IHPC), Agency for Science, Technology and Research (A*STAR),1 Fusionopolis Way, \#16-16 Connexis, Singapore 138632}

\author{Raditya Weda Bomantara}
\email{raditya.bomantara@kfupm.edu.sa}
\affiliation{Department of Physics, Interdisciplinary Research Center for Intelligent Secure Systems, King Fahd University of Petroleum and Minerals, 31261 Dhahran, Saudi Arabia}%

\begin{abstract}

{ {Discrete time crystals (DTCs) are novel out-of-equilibrium quantum states of matter which break time translational symmetry. }
{DTCs have been extensively realized in experiments, particularly their subclass that is characterized by period-doubling dynamics due to its natural occurrence in a system of periodically driven two-level, e.g., spin-1/2, particles. The realization of DTCs beyond period-doubling, including their generalizations termed discrete quasicrystals has also been made in recent years, though such experiments typically involve higher spin particles. Constructing and observing DTCs beyond period-doubling in systems of two-level particles are generally still considered an open challenge due to the latter's $\mathbb{Z}_2$ symmetry that natively only leads to period-doubling. In this work, we developed an intuitive interacting system of two-level particles (qubits) that supports the more non-trivial period-quadrupling DTCs ($4T$-DTCs).} Remarkably, by utilizing a variational algorithm, we are able to observe clear signatures of such $4T$-DTCs in a quantum processor despite the presence of considerable noise and the small number of available qubits.  {Our findings extend the landscape of time crystalline behavior by demonstrating a distinct realization of time crystallinity beyond standard period-doubling dynamics {with qubits (two-level particles)} on a NISQ-era digital quantum computer,} as well as the potential of existing noisy intermediate-scale quantum devices for simulating exotic non-equilibrium quantum states of matter.}
\end{abstract}

\maketitle

\section{Introduction}
The concept of non-ergodicity~\cite{hardy2014thermodynamics} in quantum phenomena is ubiquitous and important in quantum many-body physics~\cite{Anderson1977}. It underlies a variety of exotic physical phenomena such as eigenstate thermalization hypothesis~\cite{Deutsch1991,Srednicki1994,Yunger2023}, many-body localization~\cite{nandkishore2015many,Abanin2019}, quantum scars~\cite{Serbyn2021,Chandran2023}, quantum chaos~\cite{d2016quantum},  and time crystals~\cite{Wilczek2012,Bruno2013,Watanabe2015,Sacha2018,khemani2019brief}.In particular, the time crystals are further categorized into two classes, i.e., the continuous time crystals (CTCs)~\cite{kongkhambut2022observation,carraro2024solid,greilich2024robust,autti2021ac,Autti2022nonlinear} and discrete time crystals (DTCs). The CTCs more closely follow the original proposal of Ref.~\cite{Wilczek2012} in which continuous time-translational symmetry is spontaneously broken. The CTCs \tc{(including time quasi-crystals~\cite{AuttiVolovik2018})} have recently been implemented in various mesoscopic-scale experimental platforms~\cite{autti2021ac,Autti2022nonlinear,kongkhambut2022observation,carraro2024solid,greilich2024robust}. The second class, i.e., the DTCs\RBB{, originally proposed in}~\cite{ElseNayak2016,Khemani2016Phase}, are a type of non-ergodic phases of matter~\footnote{{Strictly speaking, a phase of matter is characterized by macroscopic behaviors that survive in the thermodynamic limit.}} that gained prominence in the recent years~\cite{Yao2017,else2020discrete,Yao2023,Liu2018clean,Liu2022discrete}.
The DTCs emerge in periodically driven systems and are characterized by the presence of an order parameter evolving at a period that is robustly locked at an integer multiple of the driving period, persisting indefinitely in the thermodynamic limit~\cite{Sacha2018,khemani2019brief,else2020discrete,Yao2023}. 


Experimentally realizing DTCs whose order parameter exhibits a much larger period than the corresponding driving period is highly desirable, as it paves the way for observing passive quantum error correction~\cite{Bomantara2021Repetition}, as well as novel dynamical physics such as Anderson localization~\cite{Delande2017,Giergiel2017} and Mott insulator transitions in the time domain~\cite{Guo2013phase,sacha2015anderson,Mierzejewski2017,GiergielSacha2018,Guo2020}.  Unfortunately, despite a considerable number of theoretical proposals for such large period DTCs \tc{in ultacold atomic systems~\cite{Sacha2015,Giergiel2020,Kuros_2020} and in spin-based systems~\cite{Surace2019,Gong2019,Poggi2022,Pizzi2021higher}}, a majority of existing experiments were only able to realize period-doubling \RBB{DTCs, such as Refs.~\cite{Mi2022, FreyRachel2022} which utilize superconducting quantum processors or Refs.~\cite{zhang2017observation,Rovny2018a,Rovny2018b,Sreejith2018,Kyprianidis2021observation} in other quantum simulation platforms,} \tc{ period-tripling~\cite{Choi2017observation} DTCs, and most recently, discrete time quasicrystals~\cite{Shinjo2024unveiling}.} Indeed, as these experiments utilize (pseudo) spin-1/2 particles {or spin-1 particles} respectively, {their underlying $\mathbb{Z}_2$ or $\mathbb{Z}_3$ symmetry only natively supports period-doubling or period tripling feature respectively.} By contrast, the theoretical proposals of Refs.~\cite{Sacha2015,Surace2019,Giergiel2020,Kuros_2020} utilize higher-spin bosonic particles, which are thus incompatible with these experiments. Nevertheless, it is worth pointing out that experimental demonstrations of large period DTCs and even their quasicrystal generalizations have recently been made in systems of higher-spin bosonic particles, such as Refs.~\cite{Kuros_2020,AuttiVolovik2018,Taheri2022}. However, to the best of our knowledge, experimental realizations of large period DTCs in systems containing strictly two-level particles are still lacking. On the other hand, two-level particles (qubits) form the most fundamental building blocks of a typical quantum processor. Meanwhile, realizing quantum phases of matter in a quantum processor has been an active research area in recent years, as it paves the way towards establishing quantum advantage that eventually leads to the development of a functional quantum computer.

{While large period DTCs in a system of two-level particles have been theoretically identified in Refs.~\cite{Poggi2022,Gong2019,Pizzi2021higher}, their detection involves either a very large number of particles, i.e., of the order of hundreds, or $n$-body interactions with $n>2$. Both of these requirements are infeasible with current technology, as the use of very large number of particles and/or $n$-body interaction results in intractable errors.} Consequently, realizing a large $n$-tupling DTC in existing systems of spin-$1/2$ particles is a highly non-trivial task. It is worth noting that a particular example of a period-quadrupling DTC was recently realized in an acoustic system~\cite{ChengZhang2022}. However, the signature of such a DTC is only observable in the boundaries of the system rather than in its bulk. Moreover, as acoustic systems are inherently classical, the obtained large period DTC may not be directly useful for the aforementioned quantum technological applications.

In this work, we develop an interacting spin-1/2 system that supports period-quadrupling DTCs (which we shall refer to as $4T$-DTCs) observable even at moderate system sizes. \RBB{Intuitively, this is made by possible by carefully designing the periodic driving scheme such that the system exhibits an emergent $\mathbb{Z}_4$ symmetry, which is required for the formation of $4T$-DTCs, even though its building blocks (two-level particles) only support $\mathbb{Z}_2$ symmetry.} The time-evolution with matrix product states (tMPS)~\cite{Vidal2004,Paeckel2019} method then enables us to numerically investigate, in a controlled manner, the effect of disorder which is necessarily present in the actual quantum devices. Indeed, existing noisy intermediate-scale quantum (NISQ) devices currently possess various kinds of noise, ranging from the relatively poor gate fidelity, the deep circuit depth, to the thermal environment noise rising from the execution of the quantum circuit~\cite{Preskill2018,Johnstun2021understanding,Lau2022nisq}. In this case, the disorder analysis we carry out in the numerical studies simulates the imperfect gate fidelity of NISQ devices. Remarkably, we found that the signatures of $4T$-DTCs are not only robust against various types of disorders, gate errors and \tc{perturbations in the initial state}, but can even be amplified in some cases. 

Finally, in the efforts towards simulating \tc{many-body condensed matter systems~\cite{smith2019simulating,google2020hartree,kim2023evidence}, including capturing the non-equilibrium quantum phenomena~\cite{Eisert2021,IppolitiKhemani2021,Mi2022,FreyRachel2022,koh2022simulation}, strongly-correlated quantum systems~\cite{RahmaniZhang2020,Pouyan2022,chen2022high} and topological quantum physics~\cite{koh2023observation,xiang2023simulating}  on superconducting quantum processors}, \RBB{particularly for realizing DTCs beyond the period-doubling class previously achieved in Refs.~\cite{Mi2022, FreyRachel2022}, } we implement our proposed system in the IBM Q quantum processor \textit{ibmq\_cairo}. Despite the various aforementioned noise occurring in our NISQ-era device, a robust period-quadrupling order parameter are clearly observed even at the accessible moderate number of qubits. \tc{Thus, our successful observation and analysis of period-quadrupling DTCs in a NISQ-era device are expected to further bridge the gap between fundamental studies of DTCs and their technological applications. In particular, achieving a large-size higher-period DTC beyond the period-doubling type \RBB{in a system of qubits} is not just a quantitative change \RBB{since generating the necessary $\mathbb{Z}_n$ symmetry from the native $\mathbb{Z}_2$ symmetry of qubits is a highly nontrivial task. Moreover, it} could lead to quantum technological applications, such as for realizing a robust quantum memory or a passive quantum error-correcting device. The generalizability of our nontrivial DTC construction based on two-level particles also provides a fertile avenue for future theoretical and experimental studies to uncover rich phenomenology in the intersection of driven quantum matter and many-body quantum simulations on a hardware.}


\section{Results}
\subsection{$4T$-Discrete Time Crystal}
\label{sec:model}
\begin{figure}[h]
	\centering
	\includegraphics[width=1.0\columnwidth,draft=false]{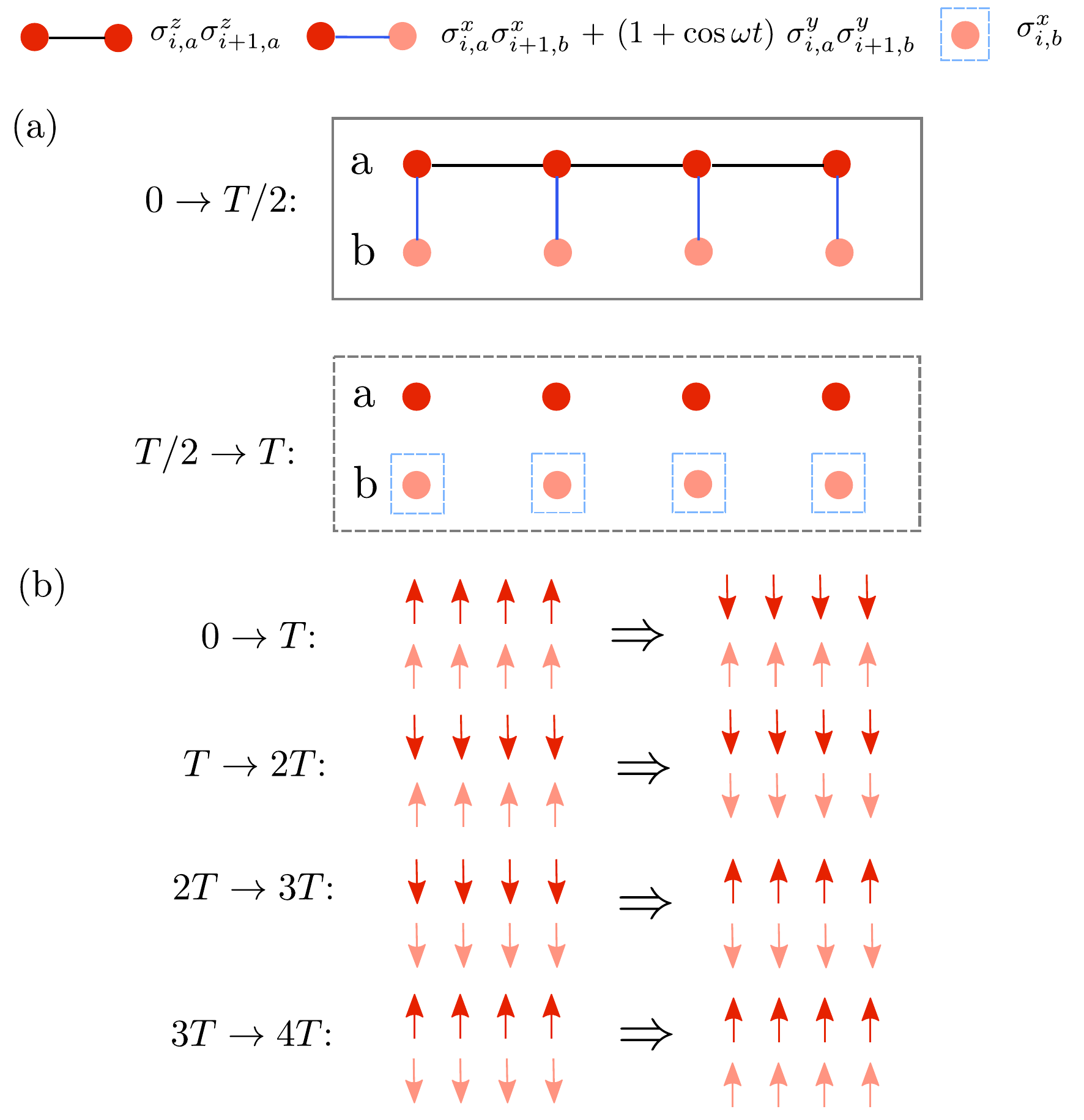}
	\caption{(a) Schematics of our periodically driven spin-1/2 ladder for $N_0=4$. During the first half of the period ($0\rightarrow T/2$, solid box), the evolution is governed by externally driven Heisenberg spin exchange interactions that are continuously modulated at frequency $\omega$. In the second half of the period ($T/2 \rightarrow T$, dashed box), the interactions are switched off and instead a magnetic field $M$ is applied in the $x$ direction . (b) The 4T-periodic oscillations can be intuitively understood in the solvable limit of $JT=0$ and $hT=MT=\pi$. With all spins initialized pointing up, the system undergoes spatially uniform 4T-periodic oscillations; ironically, these oscillations become stablized if a nonzero $JT$ is introduced.  We remark that an additional phase factors of $i$ or $(-1)$ is omitted in the illustration.
 }
	\label{fig:ModelIllustration}
\end{figure}

We propose a periodically driven spin-1/2 ladder which is schematically depicted in Fig.~\ref{fig:ModelIllustration}(a) and described by the following periodically quenched Hamiltonian,
\begin{widetext}
\begin{align}
	\label{eq:originalhamiltonian}
    \hamt = \begin{cases}
		\sum_{i=1}^{N_0} -\frac{h}{2} \left(H_i^{xx} - H_i^{yy}(t)\right)-JH_i^{zz} & \quad 0<t<\frac{T}{2}, \\
    M \sum_{i=1}^{N_0} \sigma_{i,b}^x & \quad \frac{T}{2}<t<T,
    \end{cases}
\end{align}
\end{widetext}
where $H_i^{xx}=\sigma_{i,a}^x \sigma_{i,b}^x ,H_i^{yy}(t)=\left(1+\cos{\omega t}\right)\sigma_{i,a}^y \sigma_{i,b}^y ,\tilde{H}_i^{zz}=\sum_{i=1}^{N_0-1}\sigma_{i,a}^z \sigma_{i+1,a}^z$, $\sigma_{i,a/b}^{x/y/z}$ are a set of Pauli matrices describing the spin-1/2 particle at the $i$-th site of ladder $a/b$, $N_0$ is the length of the ladder, $\omega=2\pi/T$, and $T$ is the driving period. The parameters $J$ and $h$ represent the intra- and inter-ladder interaction strength respectively, whilst $M$ describes the magnetic field strength in a spin-$1/2$ magnet analogy.  Throughout this work, we work in units such that $\hbar=1$, and set the driving period $T$ to be $1$, for easy comparison with the $4T$ time-crystal oscillation period demonstrated later. 
Note that the Floquet driving appears not just in the 2-step quench, but also in the continuous time dependence of $H^{yy}_i(t)$. In this case, the term $\cos\omega t$ in $H_i^{yy}(t)$ serves to increase the non-integrability of our system, i.e., the evolution operator over one period cannot be written as a mere product of two exponentials.


To understand how Eq.~(\ref{eq:originalhamiltonian}) has the propensity to support the sought-after $4T$-DTC, we first consider the special limit of 
$hT=MT=\pi$ and $JT=0$ (to be referred to as the solvable limit hereafter), so that the system reduces to a variation of the model introduced in Ref.~\cite{Bomantara2022}. By taking an initial state in which all spins are aligned in the $+z$-direction, which we denote as $|\uparrow \cdots \rangle_a \otimes |\uparrow \cdots \rangle_b  $, it is easily shown (using Eq.~\eqref{eq:originalhamiltonian}) to evolve as [see also Fig.~\ref{fig:ModelIllustration}(b)]
\begin{eqnarray}
    |\uparrow \cdots \rangle_a \otimes |\uparrow \cdots \rangle_b &\xrightarrow{(T)} &  \mathrm{i} |\downarrow \cdots \rangle_a \otimes |\uparrow \cdots \rangle_b  \nonumber \\
    &\xrightarrow{(2T)} & \mathrm{i} |\downarrow \cdots \rangle_a \otimes |\downarrow \cdots \rangle_b  \nonumber \\
    &\xrightarrow{(3T)} & -|\uparrow \cdots \rangle_a \otimes |\downarrow \cdots \rangle_b  \nonumber \\
    &\xrightarrow{(4T)} & -|\uparrow \cdots \rangle_a \otimes |\uparrow \cdots \rangle_b  . \label{eq:4Til}
\end{eqnarray}
That is, up to a global phase factor, the state returns to itself only after four periods. Note that if we strictly remain in the non-interacting limit $J=0$, such $4T$-periodicity will no longer hold even if the parameters $h$ and $M$ are tuned away from their special parameter values above by the slightest amount. Interestingly, by turning on the intersite interaction $J$, our results below show that the above $4T$-periodicity becomes more robust against such parameters variations. The induced robustness from the interaction of the form $\sigma_{i,a}^z \sigma_{i+1,a}^z$ could be understood from its connection to the physics of the quantum repetition codes~\cite{Bomantara2021Repetition}. {Indeed, Eq.~(\ref{eq:4Til}) suggests that any many-body Floquet eigenstate of the system takes the form of a macroscopic $4$-component cat state involving the four many-body states that are visited by the one-period evolution operator, e.g., $|\uparrow \cdots \rangle_a \otimes |\uparrow \cdots \rangle_b$, $|\downarrow \cdots \rangle_a \otimes |\uparrow \cdots \rangle_b$, $\mathrm{i} |\downarrow \cdots \rangle_a \otimes |\downarrow \cdots \rangle_b$, and $|\uparrow \cdots \rangle_a \otimes |\downarrow \cdots \rangle_b$, which is therefore a simultaneous eigenstate of all $\sigma_{i,a}^z \sigma_{i+1,a}^z$. In the language of quantum error correction, the $\sigma_{i,a}^z \sigma_{i+1,a}^z$ interaction thus serves as a set of stabilizer operators that allow the detection and removal of errors, thereby resulting in the robust structure of the many-body Floquet eigenstates at general parameter values, which are clearly long-range correlated and exhibit $\pi/2$ spectral gap responsible for generating period-quadrupling behavior under a generic initial state.} Moreover, as such an interaction renders our system truly many-body in nature, the observed robust $4T$-periodicity in the vicinity of parameters $hT=MT=\pi$ and $JT=0$ thus represents a signature of a genuine $4T$-DTC phase. In the following, we shall demonstrate the stability of this 4T-periodic behavior {and establish its DTC nature} in more detail for generic parameter values, first through a tMPS numerical simulation, {then by numerically performing finite size scaling analysis of two metrics that quantify the DTC properties of the many-body Floquet eigenstates,} and {finally by} physically simulating the system on the IBM \textit{ibmq\_cairo} quantum processor.

\subsection{Evidence of robust $4T$-DTC via tMPS}
\label{sec:disorder} Here, we employ an efficient method of time evolution with matrix product states (tMPS), where the quantum state is represented as an MPS, and the unitary time evolution operator as a matrix product operator (MPO)~\cite{pirvu2010matrix}.
To perform a tMPS study of the system, all the sites are realigned on a linear chain of length $N=2N_0$ with next-to-nearest neighbor couplings. We then implement a first-order Suzuki-Trotter algorithm with swap gates~\cite{suzuki1990,Stoudenmire2010} to carry out the time-evolution, the mathematical details of which could be found in the Methods. The illustration for the tMPS calculation of the model, numerical details, as well as the transformed Hamiltonian, is also shown in the  {Methods}.

\begin{figure}
	\centering
	\includegraphics[width=0.85\columnwidth,draft=false]{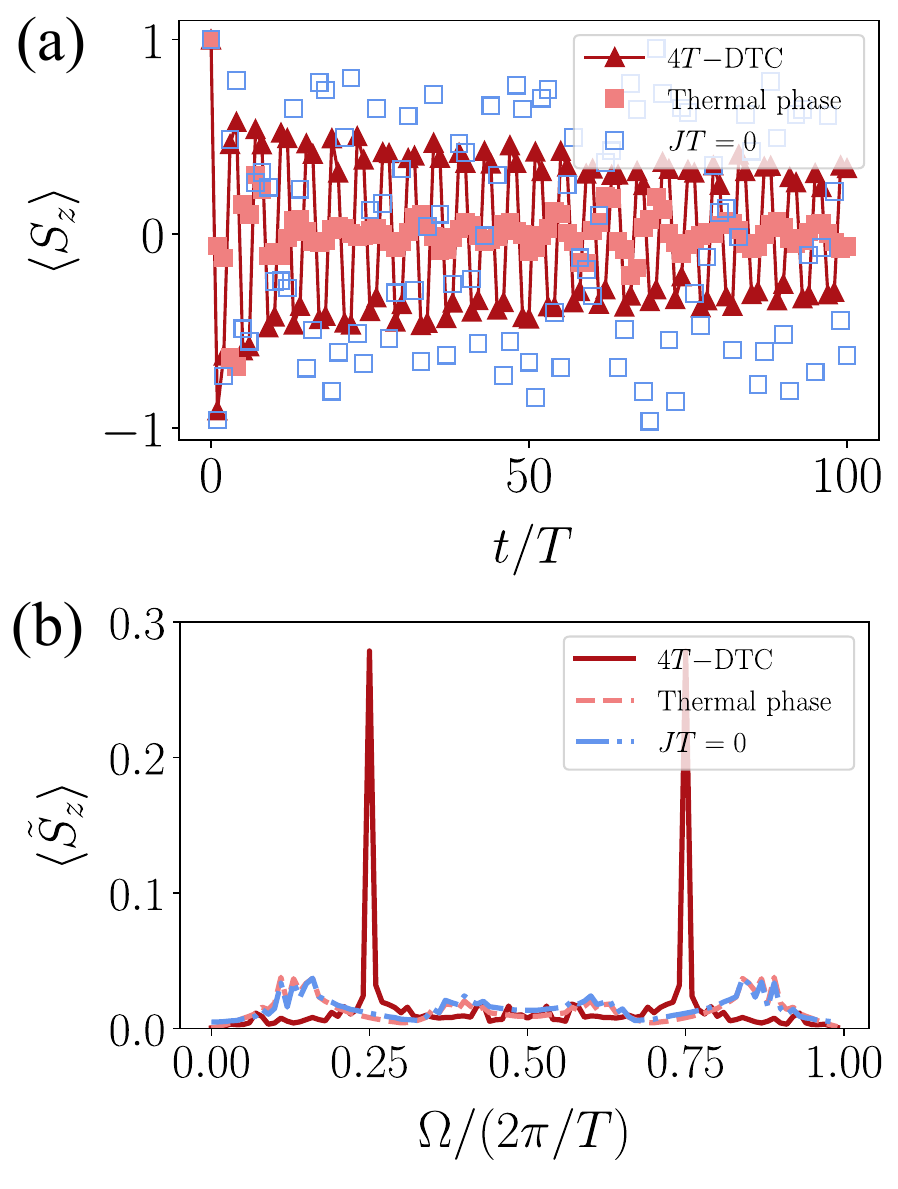}
	\caption{Numerical evidence of robust $4T$-DTC for $N=16$ sites using tMPS. (a) Magnetization $\langle S_z \rangle$ as a function of time at $MT=0.98\pi$, which is slightly perturbed away from the ``ideal" solvable limit value $MT=\pi$. {The 4T-DTC phase (red triangles) corresponds to $hT=0.9\pi$ (which is also perturbed from $\pi$) and $JT=0.16\pi$. The thermal phase (filled orange squares) corresponds to $hT=0.52\pi$ and $JT=0.1\pi$. The $JT=0$ phase (empty blue squares) corresponds to $hT=0.9\pi$.} (b) The associated stroboscopic power spectrum $\langle \tilde{S}_z \rangle$, which shows distinct frequency peaks at $\Omega = \pm \pi/2T$ only for the 4T-DTC phase. 
}

	\label{fig:N16disorderfromDTCsignature}

\end{figure}

 To capture the signatures of $4T$-DTC in our system, we calculate the stroboscopic averaged magnetization dynamics for spins residing on one of the ladders (which we choose as $a$)
\begin{align}
	\label{eq:SigmaZ}
	&\langle S_z \rangle \left(t\right) = \frac{1}{N_0}\sum_{i=1}^{N_0}\langle \sigma_{i,a}^z \rangle \left(t\right),
\end{align}
and the associated power spectrum as
\begin{align}
	\label{eq:PS}
	&\langle \tilde{S}_z \rangle \left(\Omega\right) = \left| \frac{1}{\mathcal{N}_{\rm tot}} \sum_{k=1}^{\mathcal{N}_{\rm tot}} \langle S_z \rangle\left(t\right) \exp{\left[-{\color{blue}i}\frac{k\Omega T}{\mathcal{N}_{\rm tot}} \right]} \right| 
\end{align}
where $\mathcal{N}_{\rm tot}$ is the total stroboscopic steps evolved, and $t=kT$ ($k=1,2,\cdots,\mathcal{N}_{\rm tot}$) is the stroboscopic time at step $k$. Our results are summarized in Fig.~\ref{fig:N16disorderfromDTCsignature} for two different sets of parameter values that correspond to the $4T$-DTC phase and the thermal phase respectively. Specifically, as the parameters $h$ and $M$ are chosen close to but not equal to the solvable limit values, at a finite value of the intersite interaction $J$, the {oscillatory} feature of $\langle S_z \rangle $ is clearly observed (triangle markers in Fig.~\ref{fig:N16disorderfromDTCsignature}(a)). {That this oscillatory feature indeed corresponds to period-quadrupling behavior} is further demonstrated by a sharp peak at the subharmonic frequency components $\Omega=\pi/2,3\pi/2$ in the power spectrum of Fig.~\ref{fig:N16disorderfromDTCsignature}(b). Moreover, we find that such a $4T$-periodicity is observed over a window of parameter values and not only at a specific set of parameter values, which suggests that the system indeed supports a $4T$-DTC phase. If a parameter $h$ or $M$ deviates significantly from its corresponding ideal value, $\langle S_z \rangle $ quickly decays to zero, and the system is in the thermal phase (empty square markers in Fig.~\ref{fig:N16disorderfromDTCsignature}(a)). {If the intersite interaction $J$ is absent, $\langle S_z \rangle $ may seem to exhibit finite amplitudes according to the filled square markers in Fig.~\ref{fig:N16disorderfromDTCsignature}(a). However, the power spectrum plot of Fig.~\ref{fig:N16disorderfromDTCsignature}(b) reveals that, similarly to the case of thermal phase, there are no significant peaks at any frequency, thereby demonstrating the absence of period-quadrupling behavior. Any remnant of oscillatory behavior in the cases of $JT=0$ and thermal phase as observed in Fig.~\ref{fig:N16disorderfromDTCsignature}(a) is attributed to the fact that our system Hamiltonian has the tendency to flip some ``spin" states over the course of one period, as again illustrated in the schematics of Fig.~\ref{fig:ModelIllustration}(b). }
\begin{figure}[h]
	\centering
\includegraphics[width=1.0\linewidth,draft=false]{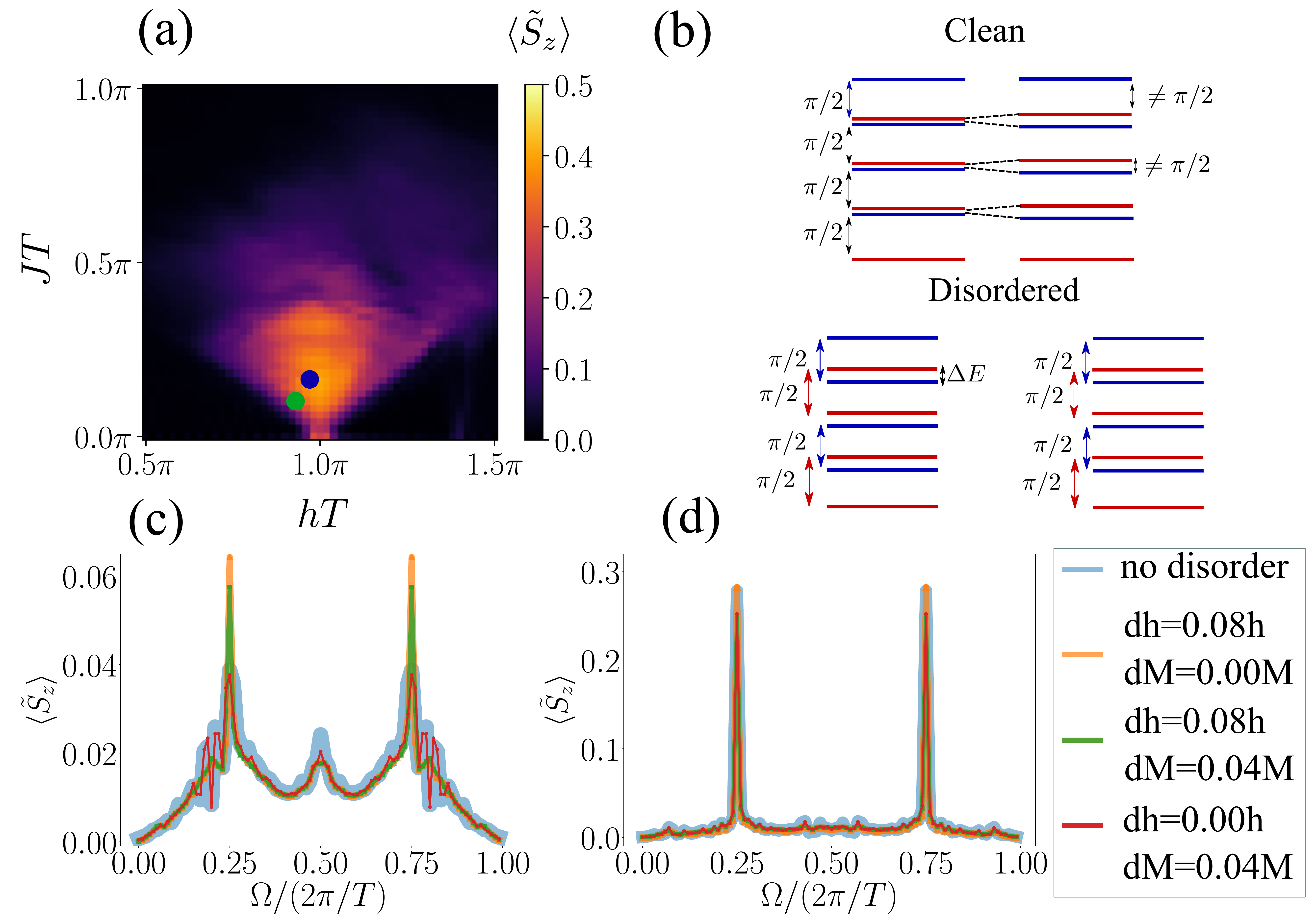}
	\caption{(a) Presence of $4T$-DTC behavior over a wide range of $JT$ and $hT$. The phase diagram representing the value of the subharmonic peak at $\Omega T=\pi/2$ and $MT=0.98 \pi$. (b) Illustration of how partial degeneracy in the clean system (overlapping blue and red lines) leads to the breakdown of the $\pi/2$ quasienergy spacing. The left (right)-side quasienergy spacings correspond to the ones before (after) perturbations. (c-d) The full power spectrum associated with the magnetization dynamics up to $t=100T$ under the influence of various disorders at (c) $JT=0.13\pi$, $hT=0.8\pi$, and $MT=0.98\pi$, i.e., green dot in panel a, and (d) $JT=0.16\pi$, $hT=0.9\pi$, and $MT=0.98\pi$, i.e., blue dot in panel a. The enhanced subharmonic peaks due to disorders are clearly observed near the DTC-thermal phase boundaries (panel c). All data points involving disorders are averaged over $220$ realizations.
 } \label{fig:disorder}
\end{figure}

In Fig.~\ref{fig:disorder}(a), we obtain the phase diagram of the {noise-free} system by plotting the subharmonic $\Omega T=\pi/2$ peak ($\langle \tilde{S}_z \rangle \left(\Omega\right)$ at $\Omega=\pi/2T$) in the power spectrum against the two system parameters $h$ and $J$. {As the value $\Omega=\pi/2T$ represents the fundamental Fourier component of a periodic variable with frequency $\pi/2T$, it is expected that a profile exhibiting a sharp peak at $\Omega=\pi/2T$ is close to being $4T$-periodic. We can then associate} a finite (zero) $\langle \tilde{S}_z \rangle \left(\Omega\right)$ at $\Omega=\pi/2T$ with the $4T$-DTC (thermal) phase{, of which the amplitude quantifies the $4T$-oscillation signature}. It is observed that the $4T$-DTC phase spans over a considerable window of $h$ values -- symmetrically about $hT=\pi$--at moderate values of $J$. At $J=0$, the period-quadrupling feature is observed only at $hT=\pi$, further confirming the role of the intersite interaction in stabilizing the DTC phase. On the other hand, at very large values of $J$, the $4T$-DTC behavior is absent altogether, which could be attributed to the presence of quantum chaos~\cite{Russomanno2017}.

In Fig.~\ref{fig:disorder}(c,d), we investigate the effect of spatial disorder on our $4T$-DTC system, {which amounts to replacing the parameters $h$, $J$, and $M$ by random values that are respectively drawn from a uniform distribution of $[P-dP +P+dP]$ for each qubit, where $P=h,J, M$ and $dP=dh,dJ, dM$. Specifically, panel (c) [(d)] is chosen so that the mean parameter values are near the border between DTC and thermal phase [deep in the DTC regime]. Such disorder in turn simulates the effect of gate imperfections in the NISQ device implementation of our 4T-DTC. As one disordered parameter might yield a different effect from another, we consider the presence of disorder on one or two parameters at a time in Fig.~\ref{fig:disorder}(c,d) for a thorough analysis. Indeed, whilst disorder in the magnetic field $M$ disorder slightly weakens the $\Omega=\pi/2T$ peak (as expected from its role as a noise), the disorder in the interaction parameter $h$ may in fact lead to an enhanced $\Omega=\pi/2T$ peak in some cases {[see Fig.~\ref{fig:disorder}(c)]}.}

{The slight enhancement of the $\Omega=\pi/2T$ peak by some disorder in some cases [see Fig.~\ref{fig:disorder}(c)] could be understood as follows.} We first recall that in a genuine $4T$-DTC, a macroscopic number of quasienergies (the eigenphases of the one-period evolution operator) form quadruplets with $\pi/2$ spacing among them, i.e., they can be written as $\varepsilon+n \pi/2$ for some $\varepsilon$ and $n=0,1,2,3$ \cite{Bomantara2021Repetition,Bomantara2022}. Ideally, such quadruplets of quasienergies should be either non-degenerate or fully degenerate ($\varepsilon+n \pi/2$ are degenerate for all $n$). In the case of partial degeneracy, i.e., $\varepsilon+n \pi/2$ are only degenerate for some $n$, certain perturbations may nonuniformly shift those degenerate quasienergies [see the upper part of Fig.~\ref{fig:disorder}(b)], which then breaks their $\pi/2$ quasienergy spacing and consequently leads to a less robust period-quadrupling signal. In the clean system, such partial degeneracy tends to occur very often; perturbing the system parameters near the DTC-thermal phase transitions then causes the many quadruplets of $\pi/2$-separated quasienergies above to break down due to the aforementioned mechanism. In the presence of spatial disorder, the system parameters for each spin or pair of spins take on slightly different values. As a result, the probability for a system's quasienergy to be degenerate is significantly reduced, thereby resulting in more robust $\pi/2$-separated quadruplets of quasienergies [see the lower part of Fig.~\ref{fig:disorder}(b)]. In {Appendix.~\ref{sec:floquetenergies}}, we further demonstrate the above argument by explicitly evaluating the quasienergy levels with and without disorder.

Away from the phase transition boundaries, the presence of disorders does not seem to yield a signal improvement. In some cases, disorders instead slightly reduce the subharmonic peak. Indeed, away from the phase transition boundaries (close to the solvable limit), the detrimental partial degeneracy among different quadruplets of $\pi/2$-separated quasienergies is already rare to begin with. In this case, disorders instead serve as perturbations in the system parameters with respect to the solvable limit values. Nevertheless, as demonstrated in Fig.~\ref{fig:disorder}(c) and (d), our DTC is remarkably robust against moderate disorders ($\sim 8\%$).

We note that in our numerical results above, we have fixed the system size at $N=16$ so that it is larger than but remains close to that accessible in our quantum processor implementation below. However, as demonstrated in {Appendix.~\ref{sec:N32results}}, qualitatively similar results are also obtained at larger $N=32$.

\subsection{{Finite size scaling analysis of the spectral properties of the many-body Floquet eigenstates}}
\label{sec:valch}

{In Sec.~\ref{sec:model}, we have analytically elucidated the origin of our 4T-DTC model in terms of the spectral structure of the many-body Floquet eigenstates, i.e., they are long-range correlated and exhibit $\pi/2$-spectral gap. These features are exact and clearly seen at ideal parameter values whereby all many-body Floquet eigenstates can be analytically shown to take the form of macroscopic $4$-component cat states. Away from these ideal values, we argued that such features due to a mechanism reminiscent of quantum error correction. To further verify this argument and therefore the validity of $4T$-DTC, we shall numerically perform a thorough finite size scaling analysis of the aforementioned spectral properties in the following.}

{To quantify the long-range correlated nature of the many-body Floquet eigenstates, we define an Ising-ZZ spin-glass (SG) order parameter as}~\cite{Khemani2016Phase,Bomantara2021Nonlocal} 

{
\begin{align}
    \label{eq: zz SG order}
    &\chi_{zz}= \mathcal{N}_{\chi}\sum_{\substack{i,i' \\ \gamma,\gamma'}}\sum_{n=1}^D\left|\langle \epsilon_n |\sigma_{i,\gamma}^z \sigma_{i',\gamma'}^z|\epsilon_n\rangle\right|^2,
\end{align}
where $\mathcal{N}_{\chi}=\frac{2(N-2)!}{N!\; D}$ is the normalization factor ($N=2N_0$ is the total system size {(number of qubits)}), $D$ is the dimension of the total system Hilbert space, $|\epsilon_n\rangle$ represents an eigenstate of the Floquet (one-period) time evolution operator governed by the Hamiltonian from Eq.~\eqref{eq:originalhamiltonian}, and $\sigma_{i,\gamma}^z$ stands for the Pauli-$z$ operator resided on ladder $\gamma=a,b$ at site $i$. There, the {first summation covers all} the indices {but excludes the cases} $i=i'$ and $\gamma=\gamma'$, i.e., all the distinctive sites within the total system.}  

\begin{figure}[h]
	\centering
\includegraphics[width=0.8\linewidth,draft=false]{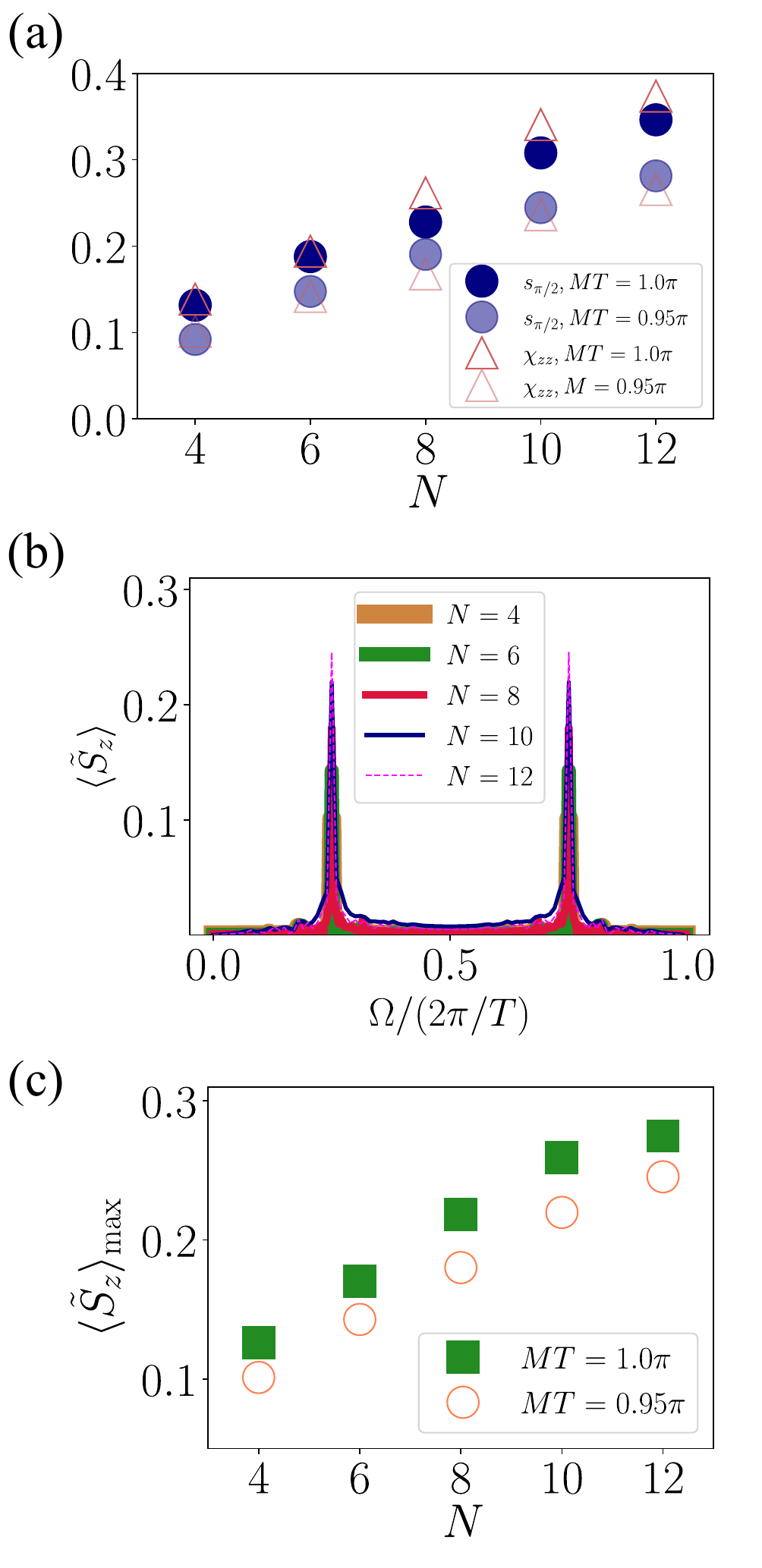}
	\caption{\tc{Finite size {scaling analysis as a} validity check of $4T$-DTC. (a) The Ising-$ZZ$ SG order $\chi_{zz}$ and {the $\pi/2$-shifted} spectral function $s_{\pi/2}$ as a function of the total system size for $M=1.0\pi$ and $MT=0.95\pi$. (b) The full power spectrum for different sizes from $N=4$ to $12$ for $MT=0.95\pi$. (c) Comparison of the averaged peak values of the power spectrum for both $MT=1.0\pi$ and $MT=0.95\pi$ as a function of system sizes. All results are obtained by exact diagonalization of the Floquet unitary operator for the time evolution of the total system up to $t=400T$. {In all panels,} other parameters used are: $JT=0.16\pi$, and $hT=0.99\pi$. }} \label{fig: sf and sg and ps}
\end{figure}

The case $\chi_{zz}=1$ signifies perfect long-range correlation with respect to $ZZ$ Pauli operator, whereas the other extreme case of $\chi_{zz}=0$ implies its absence. \tc{Our calculation, as shown in Fig.~\ref{fig: sf and sg and ps}(a) (red triangles) at generic parameters deep in the DTC region, reveals that $\chi_{zz}$ takes an intermediate value between $0$ and $1$ for $MT=1.0\pi$} \RBB{and $MT=0.95\pi$. In both cases, $\chi_{zz}$ monotonically increases with the system size, thereby demonstrating the presence of long-range correlation in the many-body Floquet eigenstates with respect to $ZZ$ Pauli operator in the thermodynamic limit. As the same signature is observed for at least two considerably different parameter values, it represents a genuine physical effect and is not a result of some fine-tuning.} 


{To quantify the tendency of the many-body Floquet eigenstates to form $\pi/2$-spectral gap, we define another useful diagnostic termed the $\pi/2$-shifted spectral function $s_{\pi/2}$, which extends the $\pi$-shifted spectral function of Refs.~\cite{Khemani2016Phase,Bomantara2021Nonlocal} for studying period-doubling DTCs. Specifically,}
{\begin{align}
    &s_{\pi/2}=\mathcal{N}_s \sum_{x \in \mathcal{M}} \int_{-\delta}^{\delta} \mathfrak{S}_{\hat{O},\pi/2}(\epsilon_n,\eta)\,d\eta,
\end{align}
where $\mathcal{N}_s=1/\sum_{n \in \mathcal{M}} \int_{-\pi/T}^{\pi/T} \mathfrak{S}_{\hat{O},\pi/2}(\epsilon_n,\eta)\,d\eta$ is the normalization factor, $\mathcal{M}$ is a set of some random integers smaller than the the total system’s Hilbert space dimension \footnote{{Throughout this work, we have chosen $\mathcal{M}=32$ for $N>4$, and $\mathcal{M}=12$ for $N=4$. }}, {the integrand is defined as} } 
{\begin{align}
    &\mathfrak{S}_{\hat{O},\pi/2}=\sum_{k=-\infty}^{+\infty}\sum_{\epsilon_m}|\langle \epsilon_n|\hat{O}|\epsilon_m\rangle|^2 \\ \nonumber &\times \delta\left[\epsilon_n-\epsilon_m-\eta-\left(2k+1/2\right)\pi/T\right] ,
\end{align}
and $\hat{O}$} is an appropriately chosen operator that is expected to map a many-body Floquet eigenstate to another that is separated in quasienergy by $\pi/(2T)$. In our calculation, we have chosen $\hat{O}=\sigma_{a,0}^z+i\sigma_{b,0}^z$ as it yields the exact desired mapping at ideal parameter values. Indeed, at ideal parameter values ($J=0$ and $hT=\pi$), $s_{\pi/2}$ can be analytically computed to yield precisely $1$, which indicates that all many-body Floquet eigenstates exhibit perfect $\pi/2$ spectral gap. At more general parameter values, our numerical calculation of Fig.~\ref{fig: sf and sg and ps}(a) \tc{(blue circles)} reveals that $s_{\pi/2}$ takes a value between $0$ and $1$, and it increases monotonically with the system size. \RBB{Similarly, as qualitatively the same signature is observed for two different parameter values $MT=\pi$ and $MT=0.95\pi$, the observed effect is not a result of fine-tuning. Instead, it genuinely demonstrates} that the many-body Floquet eigenstates indeed have the tendency to exhibit $\pi/2$-shifted spectral gap in the thermodynamic limit, and our operator choice $\hat{O}$ is relatively effective at capturing this spectral gap. 


\tc{In addition}, we have also performed a thorough analysis of the full power spectrum, calculated over a considerably long evolution time of up to $400T$, for various system sizes in \tc{Fig.~\ref{fig: sf and sg and ps}(b) for $MT=0.95\pi$ that is deviated from the perfect spin flipping parameter $MT=1.0\pi$}. As expected, the subharmonic peaks at $1/4$ and $3/4$ the driving frequency increase monotonically with the system size, thereby highlighting the lifetime improvement of the period-quadrupling behavior with the system size. \RBB{While not shown in the figure, qualitatively the same profiles are obtained for $MT=1.0\pi$}. \RBB{To clearly demonstrate this,} \tc{we have also compared the subharmonic peaks value ($\langle \tilde{S}_z \rangle_{\text{max}}$), defined as the average of peak values at $1/4$ and $3/4$, for both $MT=1.0\pi$ and $MT=0.95\pi$ in Fig.~\ref{fig: sf and sg and ps}(c). \RBB{In both cases, $\langle \tilde{S}_z \rangle_{\text{max}}$ clearly increases with the system size, thereby rigorously establishing} the lifetime improvement relation with system sizes. }

\begin{figure}[h]
	\centering
\includegraphics[width=\linewidth,draft=false]{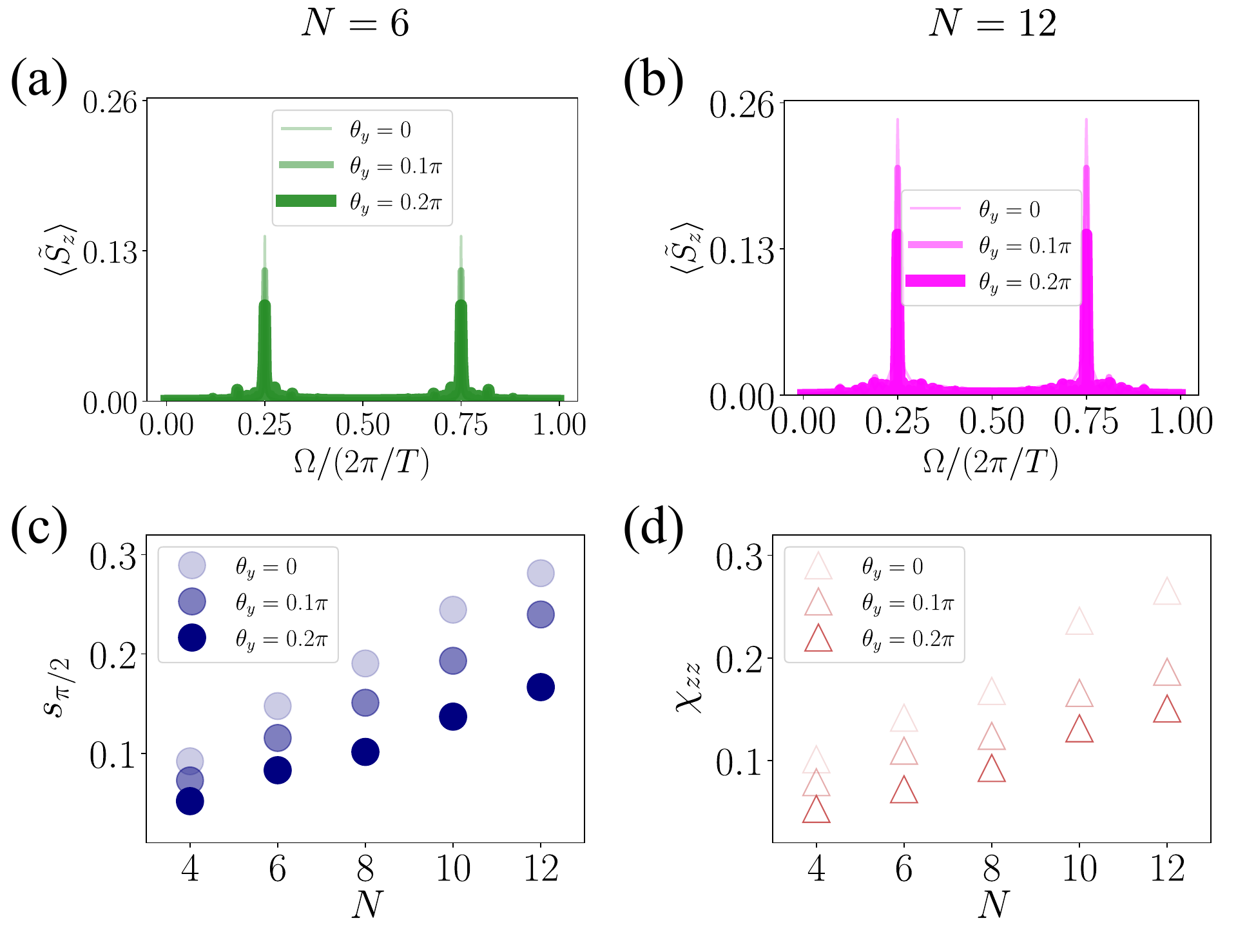}
	\caption{\RBB{Demonstration of the robustness against different initial state choices}. \tc{(a,b) the full power spectrum for different impurity initial state parameters $\theta_y$ with different system sizes in (a) $N=6$ and (b) $N=12$. (c) The $\pi/2$-shifted spectral function $s_{\pi/2}$ and (d) the Ising-$ZZ$ SG order as a function of system sizes for different impurity initial state parameters $\theta_y$.} \label{fig: robustness of the initial state}}
\end{figure}

\tc{
Finally, we have also examined the robustness of the $4T$-DTCs signatures against variations in the initial state. \RBB{This is achieved by applying an appropriate single qubit rotation with respect to Pauli $Y$ to each qubit before inputting all qubits to the circuit simulating our system. The resulting operator can be written as} $R_{y}\left(\theta_y\right)=\prod_{i} \exp\left[-i\Theta_i^y\frac{\sigma_i^y}{2}\right]$, \RBB{where the product is over all qubits involved} and $\Theta_i^y$ is randomly drawn from a uniform distribution $\left[-\theta_y, \theta_y\right]$. \RBB{In Fig.~\ref{fig: robustness of the initial state}(a,b), we repeated the calculation of power spectrum for two different system sizes and several different angular deviation $\theta_y$ values. In all cases, the subharmonic peaks at $1/4$ and $3/4$ are still significantly observed. Moreover, the larger system size $N=12$ yields larger peak values for all $\theta_y$ values under consideration, thereby demonstrating the robustness of the $4T$-oscillation against variations in the initial state, as expected from a genuine DTC.}} 

\RBB{Strictly speaking, the Ising-$ZZ$ SG parameter $\chi_{zz}$ and the $\pi/2$-shifted spectral function $s_{\pi/2}$ capture the spectral features of the system's Floquet operator and thus do not directly depend on the initial state. However, the effect of varying the initial state could be indirectly imprinted on these quantities by applying the rotating operator (via $R_y^{(-1)}(\theta_y) (\cdots) R_y(\theta_y)$) on the relevant observable, i.e., $(\cdots)\equiv \hat{\mathcal{O}}$ for $s_{\pi/2}$ and $(\cdots)\equiv \sum_{\substack{i,i' \\ \gamma,\gamma'}} \sigma_{i,\gamma}^z \sigma^z_{i',\gamma'}$ for $\chi_{zz}$. In Fig.~\ref{fig: robustness of the initial state}(c,d), several $\theta_y$-modified $\chi_{zz}$ and $s_{\pi/2}$ parameters are plotted with the system size. Remarkably, even when largely deviated from the all spin-up initial state, e.g., with $\theta_y=0.2\pi$, non-trivial values for both $s_{\pi/2}$ and $\chi_{zz}$ that improve with increasing system sizes are still clearly observed in Fig.~\ref{fig: robustness of the initial state}(c) and (d) respectively. All these results provide clear evidence that the $4T$-DTCs signatures are robust against variations from the all spin-up initial state.}

To conclude, our \tc{thorough} numerics have confirmed that our system represents a genuine $4T$-DTC based on its three defining characteristics that persist at general parameter values \RBB{and various initial states}: (i) the many-body Floquet eigenstates are long-range correlated ($\chi_{zz}$ is finite and increases monotonically with the system size), (ii) the corresponding Floquet quasienergies exhibit $\pi/2$-shifted spectral gap ($s_{\pi/2}$ is finite and increases monotonically with the system size), (iii) the period-quadrupling behavior lasts longer with increase in the system size (the subharmonic peaks at $1/4$ and $3/4$ the driving frequency increase monotonically with the system size).

\subsection{Realization of robust $4T$-DTC on a quantum processor}
\label{sec:ibmq} 
{Being a NISQ device, the currently available quantum processors inevitably exhibit various types of device noise. Among these, quantum gate imperfections,  which simulate the time-evolution of some disordered Hamiltonian, are the most significant sources of noise for simulating short- to moderate-time dynamics{~\cite{FreyRachel2022}}. That our tMPS numerical simulations of the model above have demonstrated the extremely robust signatures of $4T$-DTC in the presence of such a noise in turn suggests the feasibility for implementation on a quantum computer. }


\begin{figure}[t]
	\centering
	\includegraphics[width=1.0\linewidth]{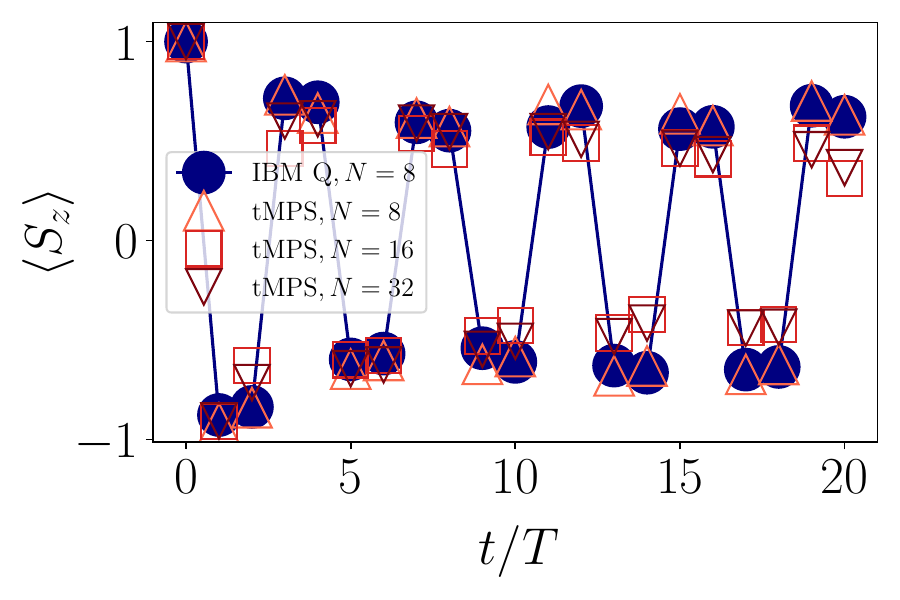}
	\caption{{The stroboscopic magnetization ($\langle S_z \rangle$) obtained from an IBM quantum processor {(solid blue circles) and tMPS at several system sizes (empty square/triangular marks)} as a function of {the number of periods ($t/T$)}.} The parameters used are: ${JT=0.16\pi}$, $MT=0.98\pi$, $hT=0.9\pi$. For the details of the device \textit{ibmq\_cairo} and its error information, see {Fig.~\ref{fig:ibmqcairoerror} in Appendix.~\ref{sec: additional results for noisy simulations}.} 
} 
	\label{fig:ibmqresults}
\end{figure}

Here, we proceed to realize our system and capture its DTC signatures on the {IBM} quantum processor. Naive implementation of the time dynamics of our model in a quantum circuit follows from a similar trotterization procedure as in our tMPS simulations, and more details are shown in the {Methods and Appendix}. For any quantum circuit implementation, the coupling between qubits needs to be implemented via basic quantum gates such as the {controlled-}NOT (CNOT) and the single-qubit rotation gates; one significant advantage of a quantum circuit implementation over other quantum platforms, e.g., ultracold atoms, is that a time-dependent model such as our DTC model can be implemented without any additional difficulty, just by concatenating different Trotter steps at different stages. Instead of transpiling local couplings within each Trotter step, we adopt a more efficient approach by leveraging the variational circuit optimization technique, and the details of this approach are presented in {Appendix.~\ref{sec:variationalalgorithm}}. This strategy of directly implementing the whole circuit requires fewer $CX$ gates and compresses the circuit depth, thereby suppressing the effect of gate error. For other simulations using the Trotterization approach~\cite{Xu2021realizing,Sims2023}, the signature of the DTC gets poorer {after every Floquet cycle} under the evolving dynamics, as the circuit depth itself grows linearly. In our unique scheme of trained circuits with a fixed circuit depth, as demonstrated below, we successfully achieve robust results for our proposed $4T$-DTC model, and in particular, it exhibits remarkably strong resilience to device noise even after many Floquet cycles. {In {Appendix.~\ref{sec: additional results for noisy simulations}}, we have also numerically simulated the system by using the direct Trotterization approach and found that a slightly smaller noise level than that in the actual NISQ device is required to observe a clear 4T-DTC signature. This in turn shows that our variational circuit optimization approach remains to be the more feasible approach to implement large-period DTCs on a present quantum computer.} 

In Fig.~\ref{fig:ibmqresults}, we present the stroboscopic magnetization $\langle S_z \rangle$ (solid circles) obtained from the IBM quantum computer over time dynamics and compare these with the numerical results (unfilled circles, squares, and triangles) obtained by the tMPS method. Remarkably, thanks to our variational method, the quantum simulation over long periods of time ($20$ Floquet steps) is possible. Within such long-time dynamics, our numerical and quantum results demonstrate an excellent agreement, indicating that our IBM Q simulation gives a perfect characterization of our $4T$-DTC model. Here, we execute our IBM Q simulation on an 8-qubit case which enables the realization of highly-compressed trained circuits for overcoming device noise \footnote{Our IBM Q simulation is scalable. Detailed discussion is shown in {Appendix.~\ref{sec:ibmqprocessorerrordetails}}.}. {This is already a sufficiently long chain for demonstrating 4T-DTC since we find that finite-size effects are qualitatively insignificant}, i.e., our tMPS results at different sizes all show qualitatively similar profiles with the IBM Q results, {both deep in the DTC regime and near the boundaries with the thermal phase (see Fig.~\ref{fig:ibmqresults}} {and {\ref{fig:finitesize} in Appendix.~\ref{sec: Finite_size effect analysis}}}). {While a strict phase of matter is formally defined in the thermodynamic limit, it is standard practice to infer phase behavior from finite-sized systems when {its key} signatures persist {over several different system sizes}. Indeed, our results, supported by both simulations {(Sec.~\ref{sec:valch})} and experiments {(Sec.~\ref{sec:ibmq})}, {clearly demonstrate} robust subharmonic response and {other expected} characteristics of the 4T-DTC phase {at varying system sizes}, indicating that the observed phenomena are not finite-size artifacts but {actual} features of a genuine non-equilibrium phase.}

\section{Discussion}
\label{sec:conclusion}
We proposed an intuitive and realistic spin-1/2 model that supports a nontrivial type of DTC, characterized by a robust period-quadrupling observable rather than the more common period-doubling type. Remarkably, we were able to explicitly capture the signatures of such $4T$-DTC both numerically and via a NISQ-era IBM quantum processor, even at relatively small number of qubits and in the presence of considerable hardware noise {(some disorder actually yields slight enhancements in the {$4T$-DTC} signatures in some cases). Remarkably, the two approaches yield excellent agreement even though the noise in the quantum processor cannot be exactly replicated in the numerical simulation.}

{Besides in the area of condensed matter systems, the physical demonstration of our $4T$-DTC system at larger system sizes on a digital quantum computer is expected to be a significant future research direction in the area of quantum computing, particularly on the topics of time crystals~\cite{Wilczek2012,Bruno2013,Watanabe2015,Sacha2018,khemani2019brief}.}
On the one hand, that our $4T$-DTC phase exists within a spin-1/2 system makes it a realistic and appropriate phenomenon for benchmarking the performance of various existing noisy intermediate-scale quantum (NISQ) devices. On the other hand, the ability to achieve a large size $4T$-DTC may also open up opportunities to harness its technological application beyond observing its subharmonic signatures, e.g., as a quantum memory or a passive quantum error correcting device. Finally, a realistic generalization of our spin-1/2 system construction that supports DTCs beyond period-quadrupling makes for a good avenue for future theoretical and experimental studies that can uncover rich phenomenology lying in the intersection of \tc{Floquet~\cite{Kyprianidis2021observation}, many-body~\cite{jotzu2014experimental,schreiber2015observation,mazurenko2017cold,salomon2019direct} or even non-Hermitian physics~\cite{ShenLee2023} {such as in ultracold atomic systems~\cite{greiner2002quantum,bloch2005ultracold,BlochZwerger2008,bloch2012quantum,MiyakeKetterle2013}}.}

\section{Methods}
\subsection{Details of Suzuki-Trotter decomposition and time evolution on IBM Q}
\label{sec:STtimeevolution}

\begin{figure}[t]
	\centering
	\includegraphics[width=1\columnwidth,draft=false]{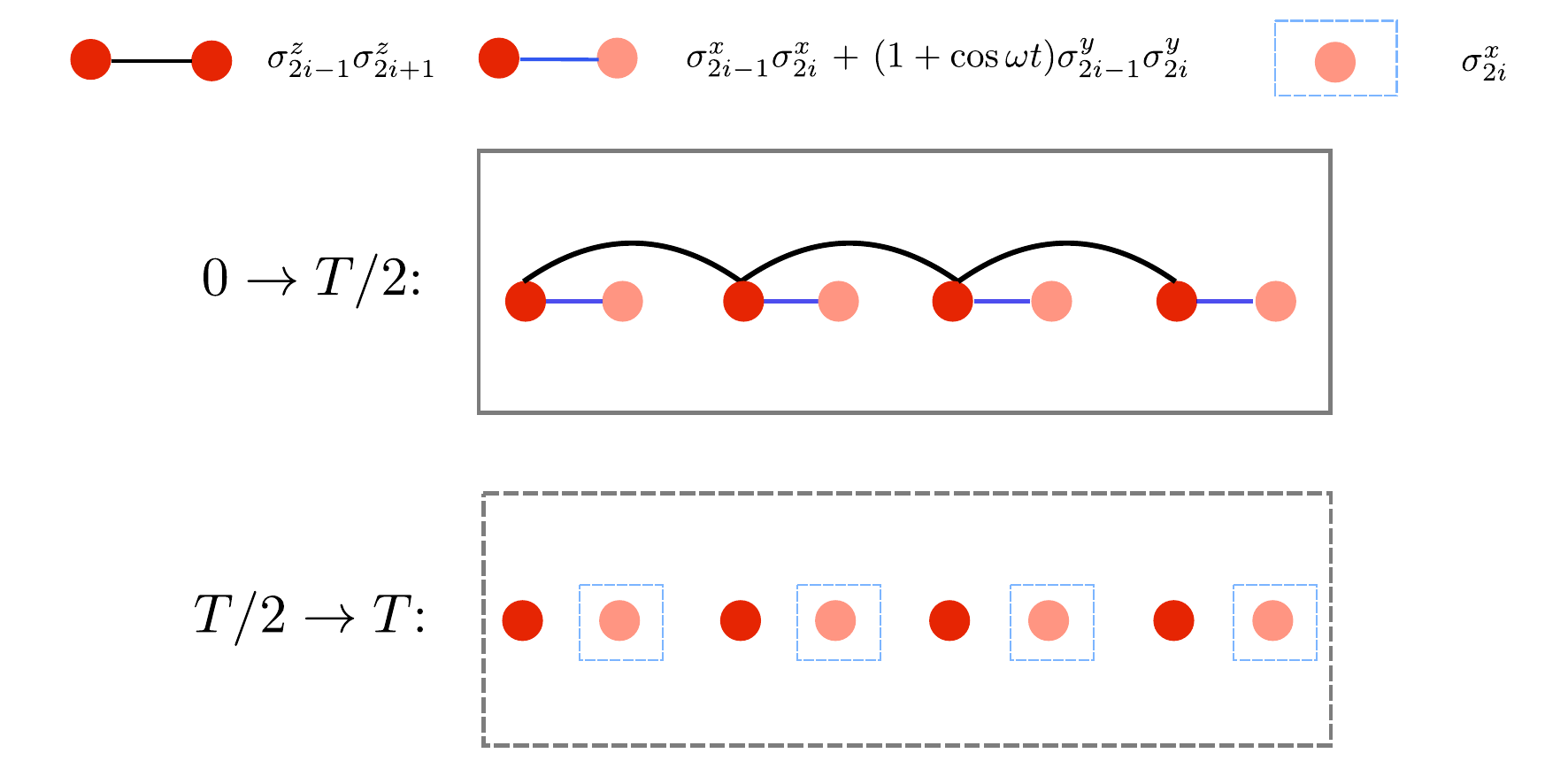}
	\caption{Illustration of schematical tMPS with Suzuki-Trotter decomposition. The total system from Fig.~1 of the main text is now mapped into a single chain with alternative nearest neighbor and next-nearest neighbor couplings.}
	\label{fig:tMPSModelIllustration}
\end{figure}

We start by discussing how to simulate the dynamics of our model in Eq.~\eqref{eq:chainhamiltonian} on a quantum circuit digitally. First, within each period $T$, the time evolution operator $\hat{U}$ operates as
\begin{align}
\label{floquetcycle}
	&\hat{U} |\psi_0\rangle= \hat{U}_{T/2 \rightarrow T}\hat{U}_{0 \rightarrow T/2} |\psi_0\rangle
\end{align}
where $|\psi_0\rangle$ is the initial state, $\hat{U}_{0 \rightarrow T/2}$ is the evolution operator for the first-half period, and $\hat{U}_{T/2 \rightarrow T}$ is the evolution operator for the second-half period. We remark that for the tMPS simulation, the total system needs to be aligned as a linear chain with some of the terms having next-nearest neighbor couplings [see Fig.~\ref{fig:tMPSModelIllustration}], and the Hamiltonian thus becomes 
\begin{equation}\label{eq:chainhamiltonian}
    \hamt = \begin{cases}
    -h/2 \sum_{i=1}^{N_0}\left(H_i^{xx} - H_i^{yy}\right)-JH_i^{zz} & \quad 0<t<\frac{T}{2}, \\
    M \sum_{i=1}^{N_0} \sigma_{2i}^x & \quad \frac{T}{2}<t<T.
    \end{cases}
\end{equation}
where 
\begin{align}
	&H_i^{xx}=\sigma_{2i-1}^x \sigma_{2i}^x \\ \nonumber 
	&H_i^{yy}=\left(1+\cos{\omega t}\right)\sigma_{2i-1}^y \sigma_{2i}^y \\ \nonumber 
	&H_i^{zz}=\sum_{i=1}^{N_0-1}\sigma_{2i-1}^z \sigma_{2i+1}^z
\end{align}

For each half period, we could therefore consider applying the first-order Suzuki-Trotter decomposition, and obtain the first half of $U$ as follows
\begin{widetext}
\begin{equation}
\label{dynamics}
\begin{aligned}
	&\hat{U}_{0 \rightarrow T/2}\approx\prod^{T/\delta t}_{n=0}\left[\prod_{i=0}^{N/4-1}e^{+i\delta tJ\sigma_{2i+2}^z \otimes I_{2i+3}\otimes \sigma_{2i+4}^z } \prod_{i=0}^{N/4}e^{+i\delta tJ\sigma_{2i}^z\otimes I_{2i+1}\otimes \sigma_{2i+2}^z }\right]\\
	&\left[\prod_{i=0}^{N/2}e^{+i\delta th/2\left(\sigma_{2i}^x \otimes\sigma_{2i+1}^x - \left(1+\cos{\omega n\delta t }\right)\sigma_{2i}^y\otimes\sigma_{2i+1}^y\right)}\right],
\end{aligned}
\end{equation}
\end{widetext}
where $\delta t$ is the discretized time step, which is set to be $0.01/T$ ($T=1$) in our numerics. Note that for the next-nearest neighbor coupling terms in the above expression, for the simulation on IBM Q, it only requires a pair of swap gates between the ancilla qubit ($i=4$) and the physical qubit ($i=6$) with system size $N=8$ {[see Fig.~\ref{fig:ibmqcairoerror} in Appendix.~\ref{sec: additional results for noisy simulations}]}, while for the simulation with tMPS, since it is aligned to a linear spin chain, for a system size of $N$ (assume $N$ is always even in this case), it requires $(N/2-1)$ pairs of the swap gates~\cite{suzuki1990,Stoudenmire2010}. Finally, the second-half time evolution operator $\hat{U}_{T/2 \rightarrow T}$ can be simply realized by the following $R_x$ rotations
\begin{align}
	&\hat{U}_{T/2 \rightarrow T}=\prod^{N/2}_{i=0}e^{-i\frac{MT}{2}\sigma_{2i+1}^x}.
\end{align}

Throughout this manuscript, we take the evolution time step for our tMPS algorithm as $0.01\hbar$, and the convergences of the tMPS calculations are confirmed by checking the truncation errors after repeating the runs for different values of the maximum bond dimension. We find that keeping a maximum auxiliary bond dimension of $100$ ($60$) for $N=24,32$ ($N=8,16$) for all tMPS calculations allows us to produce precise simulations, with errors on the observables of, at most, $10^{-8}$.

\subsection{Variational quantum algorithm for the time evolution on IBM Q}

Here, we provide the essential details of our variational quantum circuit recompilation for the time evolution operators used in the main text. The scheme of our variational algorithm to obtain the parametrized quantum circuit is depicted in {Fig.~\ref{fig:ibmqcairoerror}(c) and (d) in Appendix.~\ref{sec: additional results for noisy simulations}}. The original time evolution circuit $\hat{\mathcal{U}}$ with Suzuki-Trotter gates is transformed into a Trotterized ansatz circuit $\hat{\mathcal{V}}$ via variational optimization {[see Fig.~\ref{fig:ibmqcairoerror}(c) in Appendix.~\ref{sec: additional results for noisy simulations}]}. For each stroboscopic time step $t=kT$ ($k=1,2,\cdots,\mathcal{N}_{\rm tot}$), it is transformed into an ansatz $\hat{\mathcal{V}}(t=kT)$, which consists of an initial layer of $U_3$ gates followed by concatenated odd layers (green) and even layers (purple) {[Fig.~\ref{fig:ibmqcairoerror}(c) in Appendix.~\ref{sec: additional results for noisy simulations}]}, which has a Trotterized time evolution pattern. Here, we set all number of total layers to be equal to be three, i.e. it has in total three combined layers of odd and even layers.

The circuit variational optimization is done by minimizing the cost function 
\begin{align}
	&F(\hat{\mathcal{U}},\hat{\mathcal{V}}) = 1-\langle \psi_0 |\hat{\mathcal{V}}^{\dagger}\hat{\mathcal{U}} |\psi_0 \rangle 
\end{align}
where we have deliberately chosen the initial $|\psi_0 \rangle$ with all sites in $|\uparrow\rangle$. The process of the circuit variational optimization involves optimizing a $3D$ rotation gate labeled as $U_3(\theta, \phi, \lambda)$. This gate is characterized by three rotational parameters which are variable: $\theta$, $\phi$ and $\lambda$. We refer to the {Appendix.~\ref{sec:additional details of variational optimization}} for other details of the optimization.

\subsection{Measurement of observables on IBM Q}
The magnetization in the $z$ direction for each spin residing on ladder a from our model is obtained via the measurement procedure on IBM quantum processor, which is performed after the time evolution in the Trotterized ansatz circuit $\hat{\mathcal{V}}$. On IBM Q, the measured outcomes are all represented in binary bit strings, i.e. $0$ for spin-up ($|\uparrow\rangle$), and $1$ for spin-down ($|\downarrow \rangle$). For each site $i$, the magnetization in the $z$ direction for each spin residing on ladder a $\langle \sigma_{i,a}^z \rangle$ is computed as

\begin{align}
    \label{eq:magnetizationcalculationonibmq}
    &\langle \sigma_{i,a}^z \rangle = \langle \sigma_{i,a}^{\uparrow} \rangle -\langle \sigma_{i,a}^{\downarrow} \rangle 
\end{align}

where 
$\sigma^{\uparrow}=\begin{pmatrix}
    1 & 0 \\ 0 & 0
\end{pmatrix}$, and $\sigma^{\downarrow}=\begin{pmatrix}
    0 & 0 \\ 0 & 1
\end{pmatrix}$.

Then, the stroboscopic averaged magnetization dynamics from Eq.~\ref{eq:SigmaZ} as well as the associated power spectrum from Eq.~\ref{eq:PS} can all be easily obtained from the above results.


\section*{Acknowledgment}
We are grateful to Jiangbin Gong {and Tim Byrnes} for fruitful discussions. T.~C. thanks E.~Miles Stoudenmire for fruitful discussion {about ITensor usage} via ITensor discourse group ({https://itensor.discourse.group/}). T.~C. and R.~S. thank Truman Ng and Russell Yang for discussions on the quantum simulation implementation on IBM Quantum services. We acknowledge the use of IBM Quantum services for this work. The views expressed are those of the authors, and do not reflect the official policy or position of IBM or the IBM Quantum team. The MPS calculation in this work is performed using ITensor library~\cite{itensor}. {The computational work for this article was partially performed on resources of the National Supercomputing Centre, Singapore ({https://www.nscc.sg/}). We are also gratefully acknowledge computational resources from the A*STAR Computational Resource Center (ACRC). T.~C. also acknowledge that this work was supported in part through the NYU IT High Performance Computing resources, services, and staff expertise, made possible through the collaboration and support of Tim Byrnes.}

\section*{Funding}
C.~H.~L. and T.~C. acknowledges support by Singapore's NRF Quantum engineering grant NRF2021-QEP2-02-P09 and Singapore's MOE Tier-II grant (MOE-T2EP50222-0003). T.~C. and B.~Y. acknowledges the support from Singapore National Research Foundation (NRF) under NRF fellowship award NRF-NRFF12-2020-0005. R.~W.~B acknowledges the support provided by the Deanship of Research Oversight and Coordination (DROC) at King Fahd University of Petroleum \& Minerals (KFUPM) through project No.~EC221010. 

\section*{Author contributions}
R.~W.~B. proposed the initial idea, and developed the $4T$-DTC model. C.~H.~L. and R.~W.~.B. guided the overall
research direction and supervised this work. T.~C. performed the classical simulations using tMPS, and also developed the algorithms and implemented them
on the IBM Q. R.~S. developed part of the algorithms on IBM Q and proposed the idea of the circuit recompilation. Y.~B. contributed to part of the discussions. All the authors contributed to the writing of the manuscript.

\section*{{Data availability}}
{We have uploaded the essential codes and data which are related to this work at \cite{codeonzenodo}.}



\bibliography{refmain}

\onecolumngrid
\appendix

\section{Robustness of $\pi/2$ quasienergy spacing in the presence of disorder}
\label{sec:floquetenergies}
In the main text, we discussed that the positive effect of disorder observed near the DTC-thermal phase boundaries is attributed to the preservation of the $\pi/2$ quasienergy spacing among the quadruplets of quasienergies. In this section, we verify this argument by explicitly computing the full quasienergy spectrum of our system (via exact diagonalization) at the same system paramaters as those of Fig.~3(c) in the main text. In particular, a quasienergy $\varepsilon$ is obtained from the eigenvalue $e^{-\mathrm{i} \varepsilon  }$ of the (unitary) one-period time evolution operator. 

\begin{figure}[h]
	\centering
	\includegraphics[width=0.75\columnwidth,draft=false]{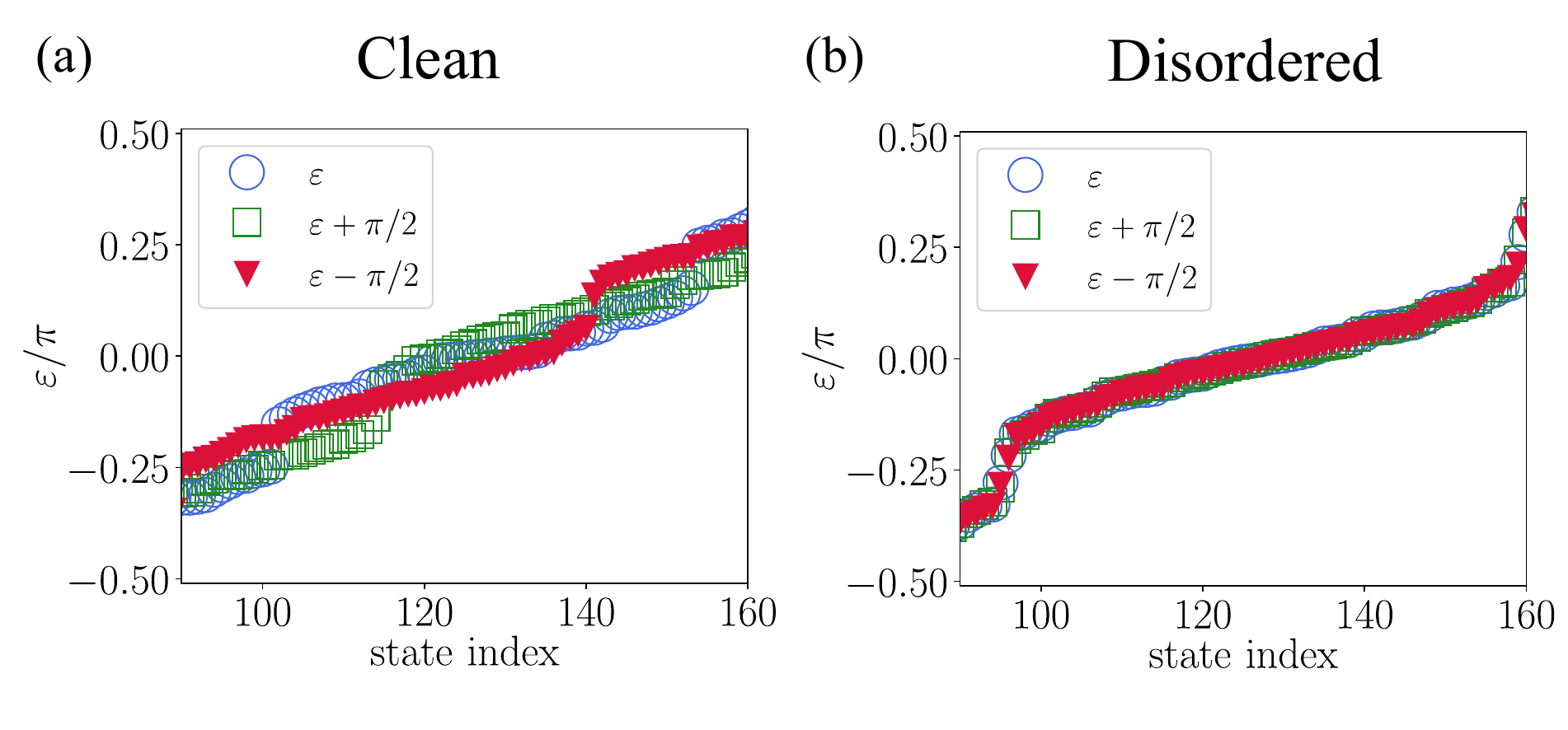}
	\caption{Representative quasienergy solutions $\varepsilon$ (blue circles) and the quasienergies that are closest to them upon shifting by $\pm \pi/2$ (see the discussion of Sec.~\ref{sec:floquetenergies} for the exact definitions) in (a) the absence of disorder and (b) the presence of spatial disorder with $dh=0.08h, dM=0.08M$. The system size is taken at $N=8$ in both panels and all other parameters used are the same as in Fig.~3(c) of the main text, i.e., $JT = 0.13\pi, hT = 0.8\pi, MT = 0.98\pi$. Notably, the degeneracy modulo $\pi/2$ is stronger when disorder is present, thereby explaining the more robust 4T-periodicity when disorder is present. } 
	\label{fig:floquetenergies}
\end{figure}

Our results are summarized in Fig.~\ref{fig:floquetenergies}, where we have shown some representative quasienergy levels of the system with and without disorders. In particular, for each data point, the blue circle marks the $j$th quasienergy solution, i.e., $\varepsilon_j$, whereas the corresponding green square (red triangle) marks some other quasienergy solution, e.g., $\varepsilon_\ell$ whose value is the closest to $\varepsilon_j +\pi/2$ ($\varepsilon_j -\pi/2$). Therefore, note that for the many-body quasienergy levels to contain a macroscopic number of $\pi/2$-quasienergy separated quadruplets, the majority of the red triangles, green squares, and blue circles in Fig.~\ref{fig:floquetenergies} must coincide with one another. As is clear from the figure, this is the case when disorder is present (panel b). In the absence of disorder (panel a), a significant number of quasienergy solutions do not coincide with any other quasienergy solution when shifted by $\pm \pi/2$. This in turn explains the stronger DTC signal observed in Fig.~3(c) of the main text when disorder is present.

\section{$4T$-DTC results for a larger system size of $N=32$}
\label{sec:N32results}

\begin{figure}[t]
	\centering
	\includegraphics[width=1.0\columnwidth,draft=false]{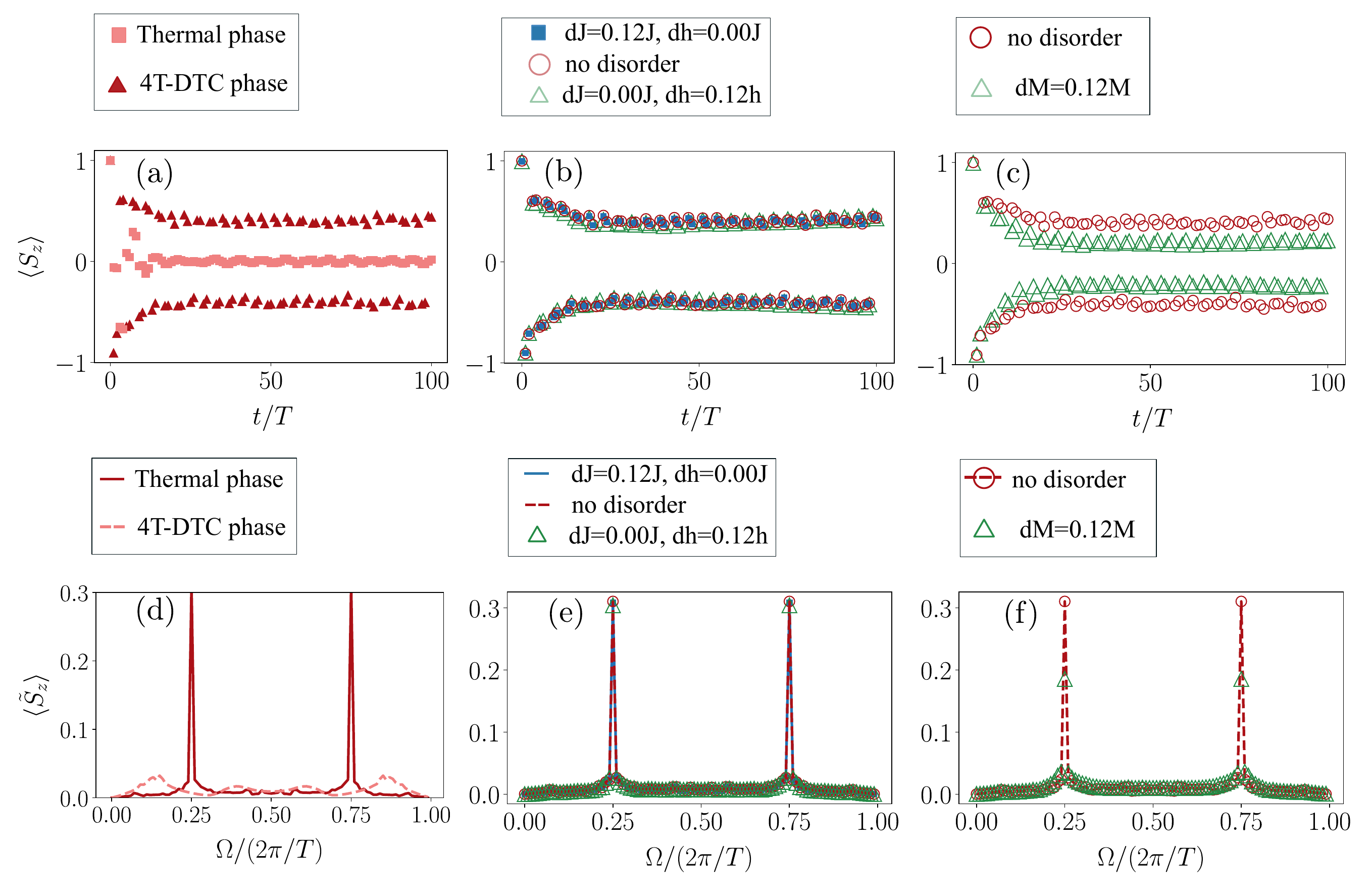}
	\caption{Numerical evidence of robust $4T$-DTC for $N = 32$ sites using tMPS. (a)-(c) Stroboscopic magnetization $\langle S_z \rangle$ as a function of time. (d)-(f) The power spectrum $\langle \tilde{S}_z \rangle$. In panel a and d, the darker red data is taken at $J=0.16\pi$, $M=0.98\pi$, and $h=0.9\pi$ (which corresponds to the $4T$-DTC phase regime in Fig.~3(a) of the main text), and the lighter red data is taken at $J=0.1\pi$, $M=0.98\pi$, and $h=0.52\pi$ (which corresponds to the thermal phase regime in Fig.~3(a) of the main text). In panel b and e, the data points correspond to no disorder (red), $d h = 0.12h$ (green), and $d J = 0.12J $ (blue). In panel c and f, the data points are taken at no disorder (red) and $d M = 0.12M $ (green). In all panels, the system size is $N=32$, $T$ is set to be $1$, and all disorder results are averaged over $220$ disorder realizations, each of which is equally drawn from the corresponding interval. Numerically, it is found that for larger system size such as $N=32$, our $4T$-DTC model is still robust against a variety of spatial disorders.
 }
	\label{fig:N32disorderfromDTCsignature}
\end{figure}

In this section, we supplement our {tMPS} results presented in Figs.~2 and 3 of the main text with those calculated at a larger system size of $N=32$. Our results, which are summarized in Figs.~\ref{fig:N32disorderfromDTCsignature} and \ref{fig:N32disorderpositive} of {the Appendix}, display qualitatively the same features as those obtained in the main text. That is, the period-quadrupling signature of the $4T$-DTC is not only robust against a variety of spatial disorders (see Fig.~\ref{fig:N32disorderfromDTCsignature}), but it may get amplified in some cases (see Fig.~\ref{fig:N32disorderpositive}). 

\section{{Finite-size effect analysis}}
\label{sec: Finite_size effect analysis}
\begin{figure}[t]
	\centering
	\includegraphics[width=1.0\columnwidth,draft=false]{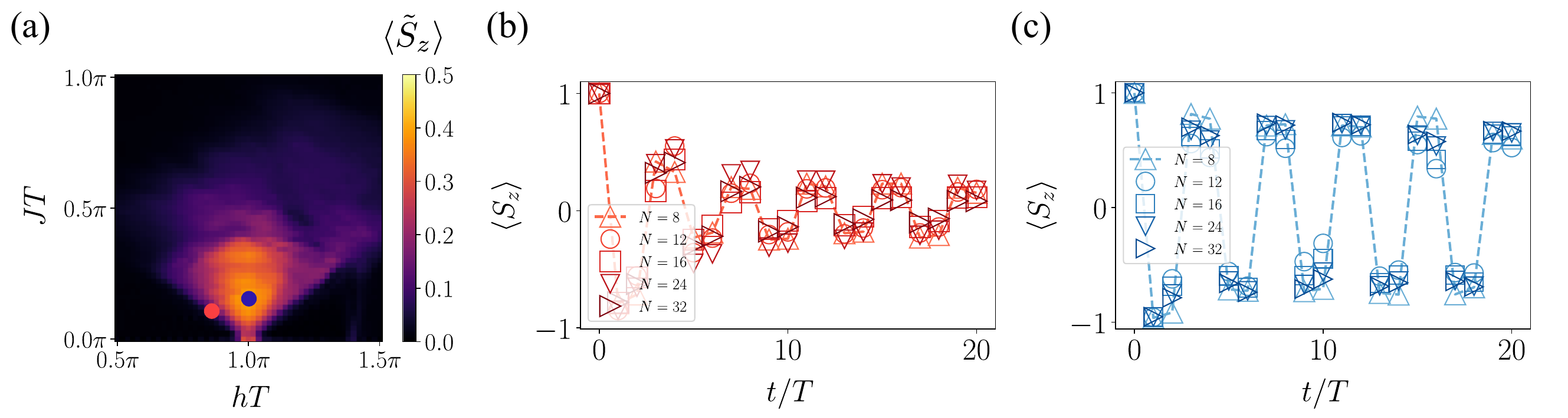}
	\caption{{Finite-size effect analysis: (a) The phase diagram representing the value of the subharmonic peak at $\Omega T = \pi/2$ and $MT = 0.98\pi$ for $N=16$ as an example. We also plotted the magnetization $\langle S_z \rangle$ as a function of time for (b) DTC-thermal phase boundary at $JT=0.14\pi,MT=0.98\pi, hT=0.8\pi$ [red dot in panel (a)], and for (c) $4T$-DTC phase at $JT=0.16\pi,MT=098\pi, hT=\pi$ [blue dot in panel (a)] both for different system size $N=8,12, 16, 24$ and $32$ using tMPS.  }}
	\label{fig:finitesize}
\end{figure}

\begin{figure}[t]
	\centering
	\includegraphics[width=1.0\linewidth,draft=false]{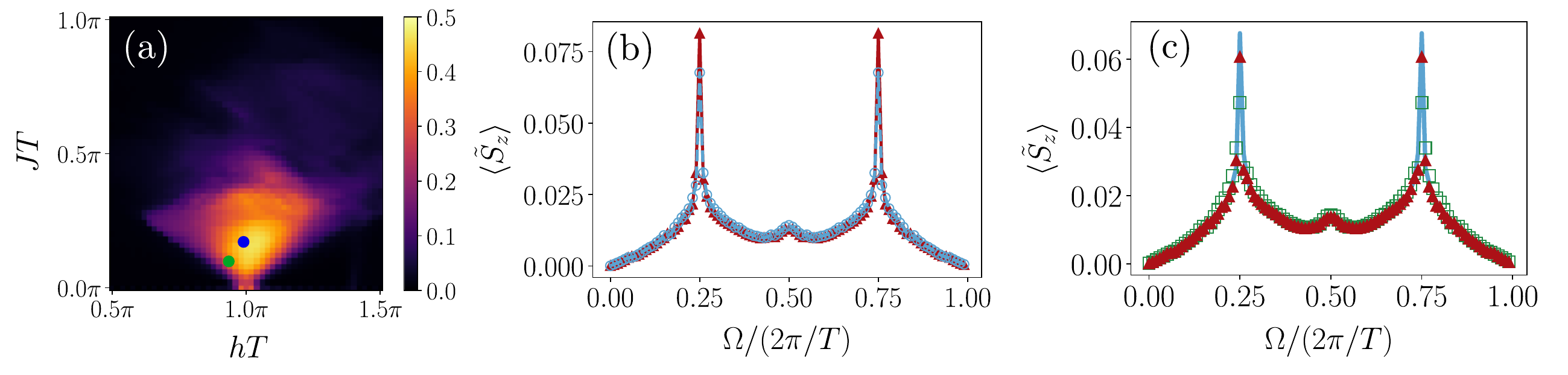}
	\caption{(a) The phase diagram representing the value of the subharmonic peak at $\Omega=\pi/2$ and $MT=0.98\pi$, numerically computed for a large $N=32$ system size. The $4T$-DTC phase is nearly symmetric, and spans over a considerable window of $h$ values, which are symmetric about $hT = \pi$ and at moderate values of $J$. The green dot corresponds to the parameter values used in (b,c), whereas the blue dot corresponds to the parameter values used in Fig.~\ref{fig:N32disorderfromDTCsignature}(b), (c), (e) and (f). (b,c) The numerically calculated power spectrum $\langle \tilde{S}_z \rangle$ over $100$ periods at $JT=0.13\pi$, $hT=0.8\pi$, $MT=0.98\pi$ with (b) no disorder (blue); $dh=0.08h$ (red). (c) without disorder (blue); only $dM=0.04M$ (green); both $dM=0.04M$ and $dh=0.08h$ (red). The system size is taken at $N=32$ in all panels and all data points involving disorders in panels b and c are averaged over $220$ disorder realizations. From the power spectrum peak in panel (b) at $\Omega T=\pi/2$, it is shown that the period-quadrupling signature of the $4T$-DTC is also enhanced by the spatial disorder $dh$. }
	\label{fig:N32disorderpositive}
\end{figure}

{For completeness, Fig.~\ref{fig:finitesize} demonstrates the finite-size effect analysis for the signatures of $4T$-DTC. Indeed, supporting the observation we made in the main text, both deep in the DTC regime and near the border with the thermal phase, the magnetization profile appears to be qualitatively the same for all system sizes. Upon closer inspection, one might observe a slight decrease in the magnetization amplitude with the system size which is vaguely visible near the border with the thermal phase and is virtually invisible deep in the DTC regime. This observation could be understood from the many-body quasienergy structure in the spirit of Fig.~3b. In particular, recall that the many-body quasienergy of a $4T$-DTC system forms quartets with $\omega/4$ separation ($\pi$/2 when taking $T=1$). Perturbations in the system parameters tend to hybridize some pairs of quasienergy eigenstates belonging to different quartets, thereby slightly shifting them in quasienergies which then leads to imperfect $\omega/4$ separation. This explains the decrease in the magnetization amplitude as one moves towards the DTC-thermal phase boundaries. As larger system sizes correspond to larger numbers of many-body quasienergy eigenstates, more hybridization events occur and consequently leads to more imperfect $\omega /4$ separation. That said, such a hybridization mechanism is generally very weak for quasienergy eigenstates that are well separated in quasienergy. These in turn explain the observed small decrease in the magnetization amplitude with the system size at fixed system parameter values. Nevertheless, as this decrease is very small and almost unnoticeable in the figure, the case $N=8$ we considered in the main text has already captured the most dominant hybridization event.}


\section{Variational algorithm for the time evolution on IBM Q}
\label{sec:variationalalgorithm}

In this section, we describe our implementation of the variational algorithm to perform the time evolution of our model in the main text on digital quantum computers, specifically the IBM Q quantum processors.

\subsection{Details of the IBM quantum processor and its error rates}
\label{sec:ibmqprocessorerrordetails}

First, we show the details of error and the choices of qubits on the $27$-qubit Falcon IBM quantum processor \textit{ibmq\_cairo}, which we have utilized throughout this work. The error profile of the device is shown Fig.~\ref{fig:ibmqcairoerror}(a), encompassing both $CX$ (CNOT) gate errors as well as readout assignment errors, as depicted by the colors in the circles (readout assignment errors) and the bonds (gate errors). In Fig.~\ref{fig:ibmqcairoerror}(b), we show the choice of $N=8+1$ qubits (grey circles) which we used to represent our system, indexed  $0$ to $8$, which is consistent with the circuit configuration in the main text.

\subsection{Additional details of the variational circuit recompilation of the time evolution operators}
\label{sec:additional details of variational optimization}
In the current NISQ-era of quantum computing, variational quantum algorithms (VQAs) have proven to be effective due to their reduced gate counts, as discussed in references~\cite{cerezo2021variational,bittel2021training}. VQAs involve a two-step process where parameterized circuits are first generated on a classical computer through an optimization algorithm. Then, these circuits with optimized parameters are executed on the quantum computer. Therefore, we explore a variational approach referred to as 'circuit recompilation' (detailed in references~\cite{heya2018variational,khatri2019quantum,SunMinnich2021,koh2022stabilizing,KohLee2022simulation}), which has shown promise in providing accurate approximations to the original unitary transformations while requiring much shorter circuit depths and fewer $CX$ and single-qubit gates compared to the default isometry decomposition. This approach results in significantly reduced overall gate errors when using current NISQ-era quantum processors~\cite{Preskill2018}.

As described in the main text, the variational circuit in this work consists of an initial layer of $U_3$ gates followed by concatenated odd layers (green) and even layers (purple)~\cite{heya2018variational,gray2018quimb,khatri2019quantum,TanSun2021,koh2022stabilizing} [Fig.~\ref{fig:ibmqcairoerror}(c)], which has a Trotterized time evolution pattern. We also follow the definitions of $3D$ rotation $U_3$ gates from Qiskit~\cite{Qiskit}:
\begin{align}
\begin{aligned}
\begin{split}U_3(\theta, \phi, \lambda) =
    \begin{pmatrix}
        \cos\left(\frac{\theta}{2}\right)          & -e^{i\lambda}\sin\left(\frac{\theta}{2}\right) \\
        e^{i\phi}\sin\left(\frac{\theta}{2}\right) & e^{i(\phi+\lambda)}\cos\left(\frac{\theta}{2}\right)
    \end{pmatrix}
    \end{split}
    \end{aligned}
\end{align}
where $\theta,\phi,\lambda \in [0,2\pi]$.

The optimization of the quantum circuits is carried out using the Limited Memory Broyden-Fletcher-Goldfarb-Shanno algorithm with box constraints (L-BFGSB), as outlined in Refs.~\cite{Malouf2002,Cao2007,koh2022stabilizing}. To prevent getting stuck in local minimums during the optimization, we employ a basin-hopping technique~\cite{li1987monte,Wales1997,Scheraga1999,wales_2004}. Here, small perturbations are introduced in each optimization iteration, which are then followed by local minimization steps.

In addition, due to the IBM quantum device configuration geometry, an additional ancilla qubit [green circle, Fig.~\ref{fig:ibmqcairoerror}(d)] is required such that the Heisenberg spin $zz$ interactions can be realized with one $CX$ gate plus two swap gates on IBM Q [Fig.~\ref{fig:ibmqcairoerror}(d)], and therefore the model configuration with $N=8$ (system) $+1$ (ancilla qubit)  is consistent with the device [Fig.~\ref{fig:ibmqcairoerror}(b)]. To simulate a larger system size, encompassing more than 10 qubits, our model described by Eq.~(1) in the main text is inherently scalable. It can be mapped to a two coupled one-dimensional chains, where each unit cell consists of two sites labeled as $a$ and $b$. In this case, the interactions between $a$ sites become next-nearest-neighbor-couplings. Such a setup can be easily achieved via the technique of circuit recompilation \cite{koh2022stabilizing}.

\subsection{Error mitigation on IBM Q}





\begin{figure}[t]
	\centering
	\includegraphics[width=1.0\columnwidth,draft=false]{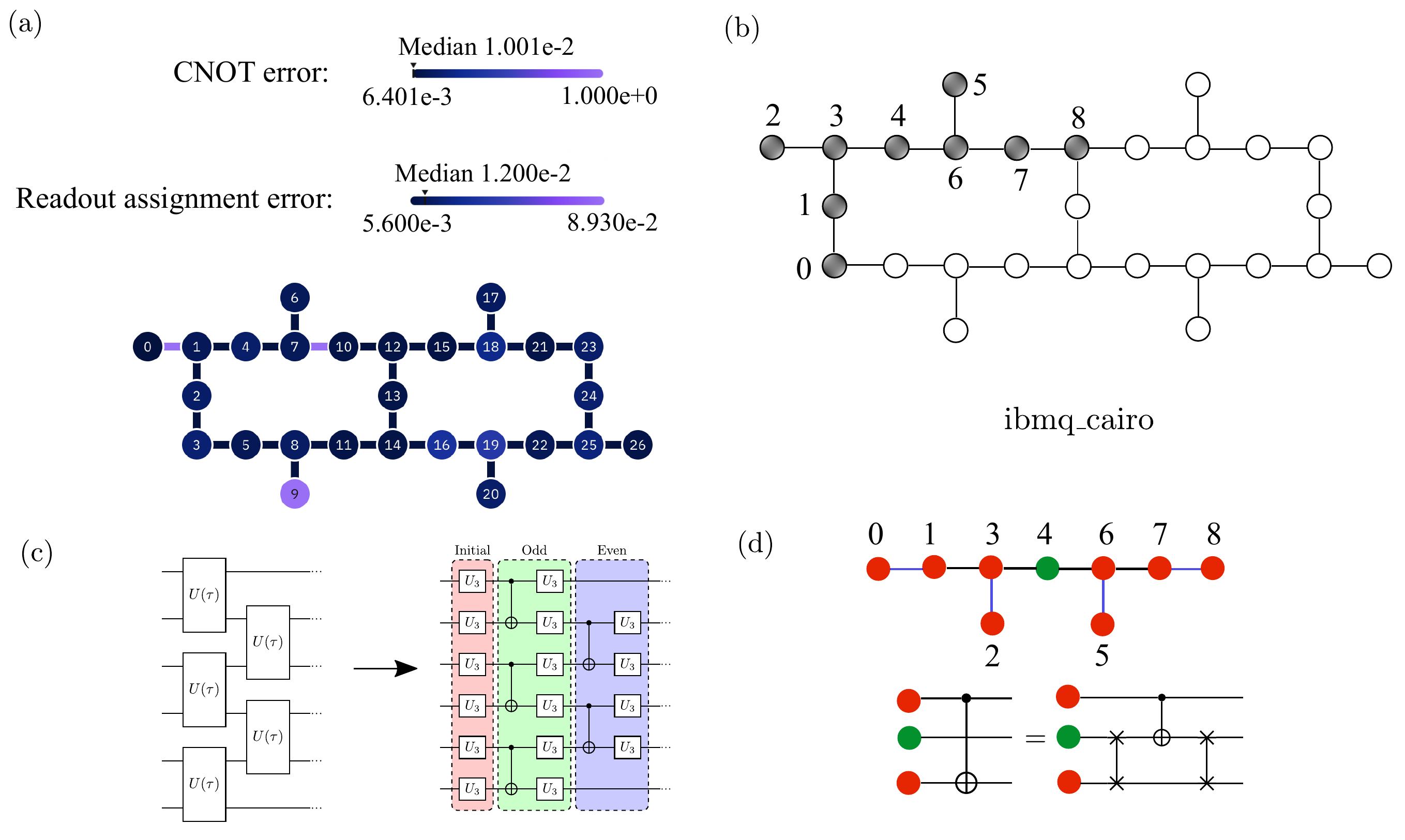}
	\caption{The geometry and error information of IBM Q device \textit{ibmq\_cairo} and details of the variational circuit recompilation algorithm. (a) The snapshot taken from IBM Q (on June 20, 2023) for CNOT ($CX$) gate error and readout error. Here, the readout error is represented by the coloring on the qubit sites, whereas the $CX$ gate error is represented by the coloring on the links connecting nearest-neighboring qubits. (b) An example of qubits choices (grey circles) of which the geometry is consistent with the configuration shown in panel (d). (c) The quantum circuit for the Trotterized time evolution is transformed into a parameterized quantum circuit suitable for execution on an IBM quantum processor via optimization on a classical computer. See also details in Ref.~\cite{heya2018variational,gray2018quimb,khatri2019quantum,TanSun2021,koh2022stabilizing} (d) (upper panel) The configuration of the quantum circuit with the additional ancilla qubit. (lower panel) The next-to-nearest neighbor $CX$ gate can be exactly decomposed to nearest-neighbor two-body gates. }
	\label{fig:ibmqcairoerror}
\end{figure}

One major issue we address in our IBM Q experiment is the readout assignment error [see also Fig.~\ref{fig:ibmqcairoerror}(a)]. This issue involves the possibility of mistakenly measuring an $|\uparrow \rangle$ state as $|\downarrow \rangle$ and vice versa. Recent advancements have made significant progress in reducing measurement errors, as documented in several studies~\cite{TemmeGambetta2017,EndoLi2018,Giurgica-Tiron2020,kandala2019error,McArdle2019,SunEndo2021,kim2023scalable}. In the context of the Qiskit environment~\cite{Qiskit}, one approach involves running calibration circuits with various initial conditions and then using the resulting data to estimate accurate measurement counts based on a calibration matrix~\cite{BravyiGambetta2021,WoottonAsfaw2021}. However, in our paper, we employ a novel readout error mitigation method~\cite{Nation2021} that requires only a small number of circuits, eliminating the need to construct a full calibration matrix, and is ready to use with their \texttt{Python} package integrated with the Qiskit environment~\cite{Qiskit}.

To align with the job submission framework of the IBM Q platform and optimize the calibration process, we combine the circuits responsible for performing the time evolution (referred to as ''physical circuits") with the calibration circuits as mentioned above into a single job submitted to the IBM Q cloud platform. This ensures that the physical circuits and calibration circuits are executed nearly simultaneously, enhancing the accuracy of the calibration process. Additionally, to maintain a consistent quantum register layout for both physical and calibration circuits, we first select and transpile the physical circuit for the specific real device using device error data calibrated by IBM Q for high-fidelity quantum nondemolition (QND) measurements~\cite{koh2022stabilizing}. We then apply this layout to the calibration circuit, ensuring that the same qubits are used for both categories of circuits. Finally, we submit both types of circuits together to the IBM Q real device for execution.

\subsection{Additional results for noisy simulations}
\label{sec: additional results for noisy simulations}
\begin{figure}
	\centering
	\includegraphics[width=0.8\linewidth]{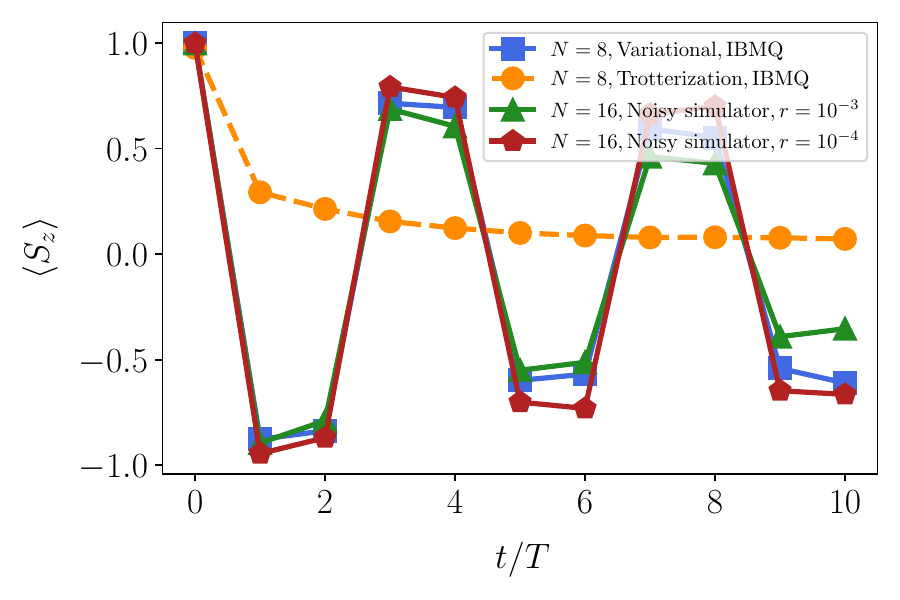}
	\caption{{Characterization of noisy effects. We present the noisy simulation on two kinds of simulators: blue and yellow curves for simulations on the real IBM Q processor; green (Trotterization approach with single-qubit gate error rate $r=10^{-3}$ shown in Fig.~\ref{fig:ibmqcairoerror} (c)) and red (Trotterization approach with single-qubit gate error rate $r=10^{-4}$) curves for simulations on the classical Aer simulator~\cite{Qiskit} with a controllable error rate $r$.}}
	\label{fig:noisyresults}
\end{figure}
{In this section, we present additional simulation results that highlight the impact of noise on the IBM Q processor, {and subsequently the critical role of our variational approach that was primarily employed in the main text. To this end, our results in this section were also obtained using the more conventional Trotterization approach, similar to what we demonstrated with our tMPS results. The comparison of these results with that obtained using the variational approach is presented in Fig.~\ref{fig:ibmqcairoerror}.} In the quantum processor used for our simulations, the average error rate for CX gates is approximately $10^{2}$, which significantly impacts the noisy simulations {based on Troterrization approach}. This is particularly evident in the comparison between the yellow and blue curves in  Fig.~\ref{fig:noisyresults}, where the noise present in the quantum processor is too large for the Trotterization approach to yield the expected $4T$-oscillation. {This clearly shows the significant advantage of our variational approach over the Trotterization approach for simulating our $4T$-DTC system with 8 qubits.}  } 

{It is worth noting that at larger system sizes, there is no guarantee anymore that our variational approach is still advantageous over the Trotterization approach, since two major challenges are at play. On the one hand, the exponential increase in complexity means that achieving good convergence in our variational approach becomes computationally expensive. On the other hand, the noise effects become even more pronounced when more qubits are involved. In particular, if noise could be reduced in future quantum processors, the Trotterization approach may become preferrable over the variational approach. In Fig.~\ref{fig:noisyresults}, we also show some results obtained from a local noise simulator under the Trotterization approach with a controllable error rate, from which we observe that {a single-qubit} error rate of around $10^{-3}$ or below is required to yield a clear $4T$-oscillation.} {Given that the latest IBM Q processor achieves a single-qubit error rate of $5\times 10^{-3}$, it is expected that a realization of our system at larger system sizes and without the use of variational approach becomes possible in the near future. However, it should be stressed that with the current technology, our variational approach is potentially the most feasible method to verify our $4T$-DTC signatures at the relatively small number of qubits. Most importantly, such an approach could also be potentially utilized to simulate other quantum many-body systems whose properties are observable at a similar number of qubits.}



\end{document}